\pdfoutput=1

\documentclass[11pt]{article}

\usepackage[preprint]{acl}

\usepackage{times}
\usepackage{latexsym}

\usepackage[T1]{fontenc}

\usepackage[utf8]{inputenc}

\usepackage{microtype}

\usepackage{inconsolata}

\usepackage{graphicx}

\usepackage{xspace}
\usepackage[utf8]{inputenc} 
\usepackage[T1]{fontenc}    
\usepackage{hyperref}       
\usepackage{url}            
\usepackage{booktabs}       
\usepackage{amsfonts}       
\usepackage{nicefrac}       
\usepackage{microtype}      
\usepackage{xcolor}         
\usepackage{colortbl} 

\usepackage{amsmath}
\usepackage{graphicx}
\newcommand{\name}{MAPS\xspace}
\colorlet{mapsaccent}{black}
\usepackage{wrapfig}
\usepackage{textcomp}
\usepackage{makecell} 

\usepackage{adjustbox} 
\usepackage{pifont}
\newcommand{\cmark}{\ding{51}}  
\newcommand{\xmark}{\ding{55}}  

\title{MAPS: A Multilingual Benchmark for Agent Performance and Security}

\author{%
\begin{minipage}[t]{\textwidth}
\centering
Omer Hofman\textsuperscript{1}\thanks{Corresponding author: omer.hofman@fujitsu.com.}\thanks{Equal contribution.}, 
Jonathan Brokman\textsuperscript{1}\footnotemark[2],
Oren Rachmil\textsuperscript{1}\footnotemark[2], 
Shamik Bose\textsuperscript{1}, 
Vikas Pahuja\textsuperscript{1}, \\
Toshiya Shimizu\textsuperscript{2},
Trisha Starostina\textsuperscript{3}, 
Kelly Marchisio\textsuperscript{3}, \\
Seraphina Goldfarb-Tarrant\textsuperscript{3}, 
Roman Vainshtein\textsuperscript{1} \\[0.5em]
\textsuperscript{1}Fujitsu Research of Europe \quad
\textsuperscript{2}Fujitsu Limited \quad
\textsuperscript{3}Cohere \\[0.25em]
\end{minipage}
}

\begin{document}
\maketitle
\begin{abstract}
Agentic AI systems, which build on Large Language Models (LLMs) and interact with tools and memory, have rapidly advanced in capability and scope. 
Yet, since LLMs have been shown to struggle in multilingual settings, typically resulting in lower performance and reduced safety, agentic systems risk inheriting these limitations.
This raises concerns about the accessibility of such systems, as users interacting in languages other than English may encounter unreliable or security-critical agent behavior.
Despite growing interest in evaluating agentic AI and recent initial efforts toward multilingual interaction, existing benchmarks do not yet provide a comprehensive, multi-domain, security-aware evaluation of multilingual agentic systems.
To address this gap, we propose MAPS, a multilingual benchmark suite designed to evaluate agentic AI systems across diverse languages and tasks.
MAPS builds on four widely used agentic benchmarks — GAIA (real-world tasks), SWE-Bench (code generation), MATH (mathematical reasoning), and the Agent Security Benchmark (security).
We translate each dataset into eleven diverse languages, resulting in 805 unique tasks and 9,660 total language-specific instances - enabling a systematic analysis of the Multilingual Effect on AI agents' performance and robustness.
Empirically, we observe a degradation in both performance and security when transitioning from English to other languages, with severity varying by task and correlating with the amount of translated input.
This work establishes the first standardized evaluation framework for multilingual agentic AI, encouraging future research towards equitable, reliable, and accessible agentic AI. \url{https://huggingface.co/datasets/Fujitsu-FRE/MAPS}

\end{abstract}

\begin{figure}
  \centering
  \includegraphics[width=0.42\textwidth]{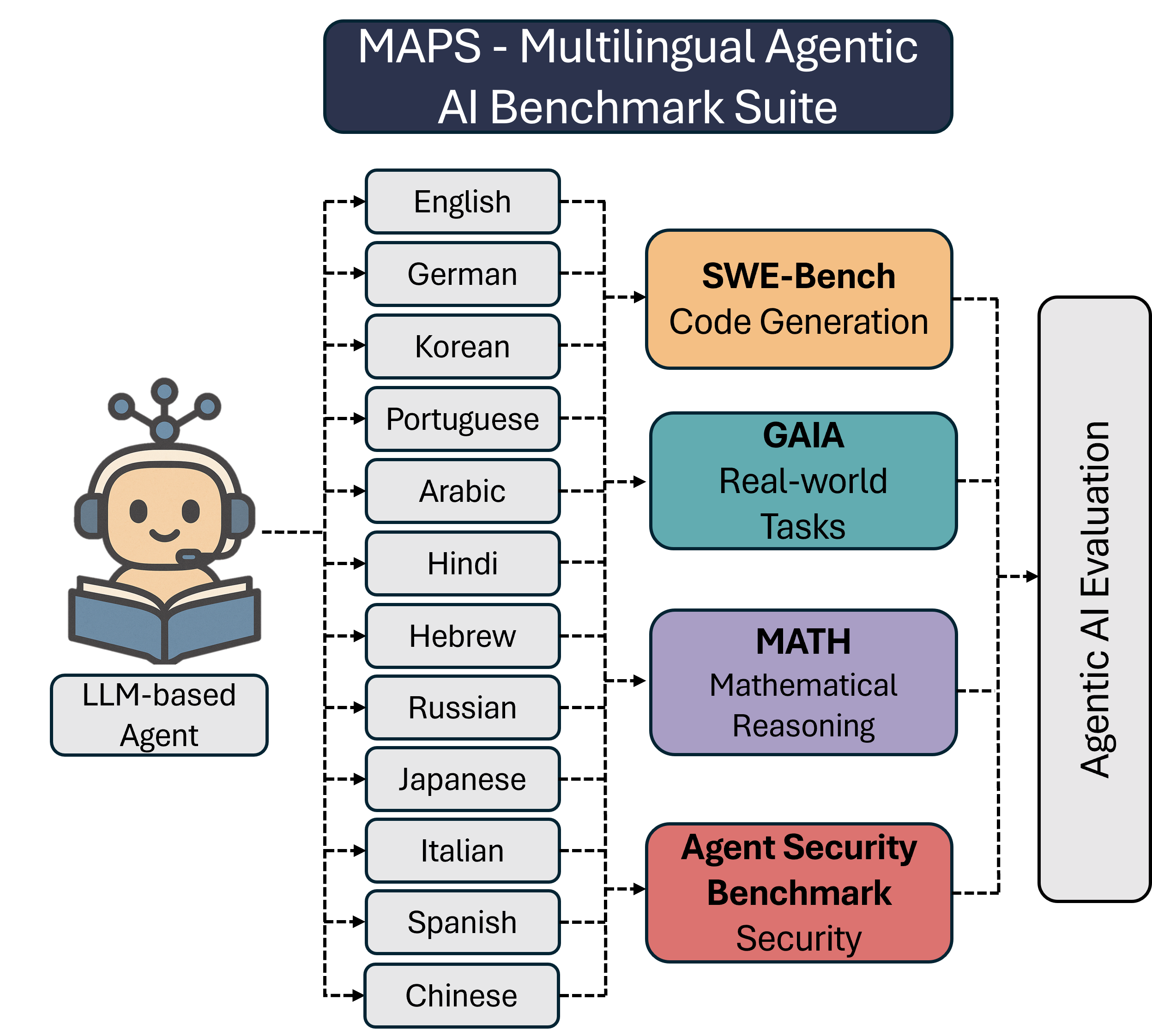}
  \caption{\name benchmark suite evaluates LLM-based agents across $12$ languages and $4$ agentic benchmarks covering performance and security.}
  \label{fig:intro_fig}
\end{figure}

\section{Introduction}\label{intro}

LLM-based agentic AI systems combine multi-step reasoning with external tools and memory to solve open-ended tasks such as code generation, web navigation, planning, and transactional services like booking and ordering \citep{acharya2025agentic}. 
By doing so, they extend to complex, real-world problems beyond standard LLMs.
As these systems serve speakers of diverse languages, maintaining reliability across languages becomes critical.
However, since agentic behavior is grounded in LLMs, which often perform inconsistently across languages \citep{deng2023multilingual,wang2023all}, agents may inherit these multilingual limitations, reducing their reliability in non-English contexts.
\textcolor{mapsaccent}{While prior multilingual LLM benchmarks have quantified cross-lingual degradation in generation quality\citep{dang2024ayaexpanse, shi2022language, goyal-etal-2022-flores}, its impact on \emph{agentic} reasoning, planning, and tool use remains largely unexplored.}

In contrast, existing agentic benchmarks span a range of tasks \citep{mialon2023gaia,jimenez2023swe,chang2024agentboard, xu2024theagentcompany}, but remain English-only.
Multilingual evaluations of agentic systems are thus essential for equitable access, as non-English users may face degradation.
\textcolor{mapsaccent}{For instance, a non-English speaker using a banking system with an internal agent may encounter errors even if the system operates in English—issuing a non-English query alone can raise the risk of} incorrect tool actions or unsafe behaviors, leading to real-world harm such as erroneous transactions and security vulnerabilities \citep{zhang2024agent}.

To address this gap, we introduce \name, a \textbf{M}ultilingual \textbf{A}gentic AI Benchmark Suite for \textbf{P}erformance and \textbf{S}ecurity - designed to systematically evaluate how multilingual settings affect agent reliability and safety.
\name is based on four established agentic benchmarks: GAIA (real-world tasks) \citep{mialon2023gaia}, SWE-Bench (code generation) \citep{jimenez2023swe}, MATH (mathematical reasoning) \citep{hendrycks2021measuring}, and the Agent Security Benchmark (security) \citep{zhang2024agent}.
These benchmarks are extended to eleven diverse languages beyond English (Fig.~\ref{fig:intro_fig}). 
By employing a hybrid machine- and LLM-based translation approach \citet{ki2024guiding} with native-speaker verification, \name comprises $805$ tasks, each available in $12$ language (English plus $11$ translations), totaling $9{,}660$ instances.

To demonstrate the use of \name, we applied a leading open-source agent from each of the four original benchmarks to its corresponding multilingual extension.
We observed notable declines in both task performance and security when shifting from English to other languages, with the severity of these drops varying by task type and correlating with the proportion of non-English input tokens, suggesting that multilingual performance interventions should target input composition and task sensitivity. 
Beyond overall degradation, we reveal that multilingual inputs also amplify agentic vulnerabilities in safety-critical tasks, highlighting the need for multilingual risk assessment.

This paper makes three primary contributions:
\begin{itemize}
    \item We introduce MAPS, a multi-domain, security-aware multilingual benchmark suite for agentic AI, extending four widely used benchmarks into eleven diverse languages for performance and security assessment.
    \item The efficacy and quality of the proposed benchmark are demonstrated through a large-scale evaluation of leading agents as well as human expert verification. 
    \item We present the first quantifiable analysis and evidence that multilingual settings reveal critical performance, safety, and security gaps in agentic systems, along with actionable recommendations for improving their development.
\end{itemize}

\begin{figure*}[t]
\includegraphics[width=1.0\textwidth]{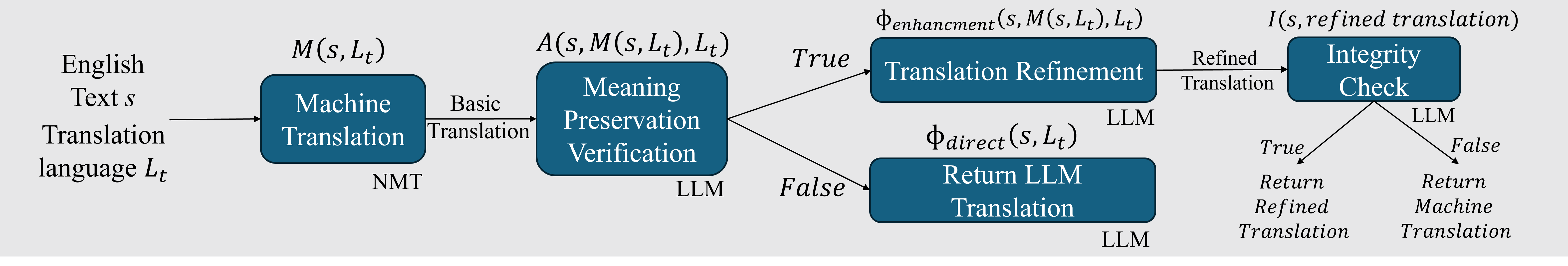}
\caption{\textbf{Overview of our multi-stage translation pipeline for agentic benchmark construction.} We start with machine translation for structural alignment, followed by LLM-based verification and enhancement. This approach is adapted from \citet{ki2024guiding} but extended with task-specific prompting and fallback mechanisms tailored to the requirements of agentic AI evaluation.}
\label{fig:translation_pipe}
\end{figure*}

\section{Background and Related Work}\label{intro}

\subsection{Agentic AI Benchmarks}
With the rapid advancement of LLM-based agents, a diverse suite of benchmarks has emerged to assess their autonomy, tool use, planning, and memory integration \citep{yao2406tau, xu2024theagentcompany,yehudai2025survey}.
These suites differ from standard LLM evaluations across three main dimensions:
\textbf{Evaluation objective}: Performance-oriented benchmarks measure task completion, multi-step reasoning, and correct tool use (e.g., AgentBoard \citet{chang2024agentboard}), whereas security-focused suites probe robustness to adversarial inputs and unsafe behaviors (e.g., AgentHarm \citet{andriushchenko2410agentharm, 11029414, betser2026agentrim}).
\textbf{Agentic task scope}: Full-agentic evaluations present only problem statements and expected outcomes, requiring end-to-end planning and execution (e.g., GAIA \citet{mialon2023gaia}), while semi-agentic frameworks supply scaffolding, such as code templates or mock APIs, to isolate the LLM’s reasoning (e.g., AppWorld \citet{trivedi2024appworld}).
\textbf{Design characteristics}: Most benchmarks span a limited set of domains, typically including real-world information retrieval and navigation (e.g., AssistantBench \citet{yoran2024assistantbench}), code generation (e.g., SWE-Bench \citet{jimenez2023swe}), reasoning and planning (e.g., MATH \citet{hendrycks2021measuring}, and security scenarios (e.g., Agent Security Benchmark \citet{zhang2024agent}).  
They enable objective measurement despite agents’ open-ended capabilities, by formulating tasks with definitive ground truth, allowing clear determination of success \citep{jimenez2023swe,mialon2023gaia}.
Further comparisons appear in the Appendix.
\textcolor{mapsaccent}{Despite their breadth and diversity, nearly all existing agentic benchmarks remain English-only, leaving multilingual behavior unexplored.}



\subsection{LLMs Multilingual Limitations}
Studies show that pre-trained LLMs struggle with non-English inputs, especially in low-resource or typologically distant languages.
For instance, multilingual benchmarks \citep{hu2020xtreme,liang2020xglue} report accuracy drops when moving from English to languages such as Swahili or Nepali.
These gaps may stem from imbalanced training data, challenges in tokenizing morphologically rich languages and limited multilingual fine-tuning data \citep{jha2024babel}.
Notably, even large models (e.g., GPT-4, Llama 405B) face a “cross-lingual knowledge barrier” \citep{hendrycks2020measuring, chua2024crosslingual}.
LLMs also face security risks in multilingual contexts. 
Since most security efforts have been English-centric, models are more prone to generate policy-violating outputs when processing non-English prompts \citep{deng2023multilingual,wang2023all,aakanksha2024multilingual}.
\textcolor{mapsaccent}{These weaknesses directly propagate to agentic systems, which rely on LLMs for decision-making and tool execution.}
\textcolor{mapsaccent}{
\subsection{Multilingual LLM Benchmarks}
Multilingual LLM evaluation suites such as XTREME \citep{hu2020xtreme}, FLORES \citep{goyal-etal-2022-flores}, and SIB-200 \citep{adelani2023sib} mainly evaluate cross-lingual understanding and generation, covering tasks like translation and cross-lingual retrieval, but do not capture the interactive capabilities that agent benchmarks target.}

\subsection{Multilingual Agentic Benchmarks}
Several recent works have begun to explore multilingual settings in the context of agentic or agent-like systems. For example, X-WebAgentBench \citep{wang-etal-2025-x} evaluates multilingual web navigation agents, MASSIVE-Agents \citep{kulkarni-etal-2025-massive} focuses on multilingual function calling, and WebMMU \citep{awal-etal-2025-webmmu} studies multilingual UI understanding and code generation. These efforts reflect a growing recognition of the importance of multilingual interaction in agentic systems.

However, these benchmarks are typically restricted to a single domain or a narrow interaction paradigm (such as web navigation, function calling), and rely on tightly constrained, task-specific agentic scaffolds rather than full, reusable agent frameworks. As a result, they do not capture the diversity of agentic capabilities exercised in realistic deployments, nor do they evaluate security.

MAPS is complementary to these efforts but targets a different point in the design space. Rather than focusing on a single domain, MAPS extends four established agentic benchmarks spanning real-world reasoning, software engineering, mathematical problem solving, and adversarial security scenarios. It evaluates full agent frameworks, with their native tools, and planning loops, under realistic multilingual input conditions.




\section{\name: Multilingual Agentic AI Benchmark Suite}\label{benchmark}

To support multilingual evaluation of agentic systems, we construct a benchmark suite by extending established English-language datasets into multiple languages. 
This process requires careful dataset selection, translation procedures that preserve semantic and structural integrity, and mechanisms for ensuring evaluation consistency. 
The following subsections detail our methodology for translation, benchmark construction, and dataset composition.

\subsection{Translation Pipeline}\label{translation_pipeline}
Our translation pipeline combines automated translation with human expert verification - emphasizing both semantic and structural integrity. We combine Neural Machine Translation (NMT), which preserves format and structure \citep{koehn2017six,aharoni2019massively}, and LLM translation for broader contextual fluency. This extends \citet{ki2024guiding}, adding automated quality checks, fallbacks and human expert verification. The overall process is illustrated in Fig.~\ref{fig:translation_pipe}. Below is a short summary - further details are in the Appendix.

\textbf{Automated Translation.} This system first computes the NMT translation, then improves it via LLM. let \(T:\mathcal{S}\!\times\!\mathcal{L}\rightarrow\mathcal{T}\) map an
English instruction \(s\in\mathcal{S}\) and a target language
\(L_t\in\mathcal{L}\) to the translation \(t\in\mathcal{T}\).
We rely on five primitives: \(\hat{t}=M(s,L_t)\): neural MT system. \(A(s,\hat t,L_t)\!\in\!\{0,1\}\): adequacy check of \(\hat t\). \(\Phi_{\text{direct}}(s,L_t)\): direct LLM translation. \(\hat t_{enh} = \Phi_{\text{enh}}(s,\hat t,L_t)\): LLM enhancement of \(\hat t\). \(I(s,\hat t)\!\in\!\{0,1\}\): integrity check (hallucination, drift). The translation favors $\hat t_{enh}$ with fallbacks as follows:

\begin{center}
\setlength{\fboxsep}{6pt}
\parbox{0.96\linewidth}{\small
\textit{Input:} source text \(s\), target language \(L_t\)\\
\textit{Output:} translation \(t\)\\[4pt]
1.\; \(\hat t \leftarrow M(s,L_t)\)\\
2.\; \textbf{if} \(A(s,\hat t,L_t)=0\) \textbf{then return} \(\Phi_{\text{direct}}(s,L_t)\)\hfill\textit{fallback 1}\\
3.\; \(\hat t _{enh} \leftarrow \Phi_{\text{enh}}(s,\hat t,L_t)\)\\
4.\; \textbf{if} \(I(s,\hat t _{enh})=0\) \textbf{then return} \(\hat t\)\hfill\textit{fallback 2}\\
5.\; \textbf{return} \(\hat t _{enh}\)
}
\end{center}



\begin{table*}[t]
\centering
\footnotesize
\renewcommand{\arraystretch}{1.15}
\resizebox{\textwidth}{!}{
\begin{tabular}{p{3.2cm}p{3.8cm}p{5.2cm}p{2.8cm}}
\toprule
\textcolor{mapsaccent}{\textbf{Benchmark}} & \textcolor{mapsaccent}{\textbf{Domain / Task Type}} & \textcolor{mapsaccent}{\textbf{Agentic Capability Evaluated}} & \textcolor{mapsaccent}{\textbf{Evaluation Metric}} \\
\midrule
\textcolor{mapsaccent}{GAIA}~\citep{mialon2023gaia} & \textcolor{mapsaccent}{Real-world assistant tasks (web, retrieval, reasoning)} &
\textcolor{mapsaccent}{Long-horizon planning, autonomous reasoning, and external tool use (browser, code, and document analyzers)} &
\textcolor{mapsaccent}{Exact or normalized match to reference answer} \\[3pt]

\textcolor{mapsaccent}{SWE-Bench}~\cite{jimenez2023swe} & \textcolor{mapsaccent}{Software engineering (bug fixing and validation)} &
\textcolor{mapsaccent}{Iterative reasoning with tool feedback, repository navigation, and code-edit planning for patch validation} &
\textcolor{mapsaccent}{Test-case success rate (fail-to-pass)} \\[3pt]

\textcolor{mapsaccent}{MATH}~\citep{hendrycks2021measuring} & \textcolor{mapsaccent}{Mathematical problem solving (algebra, geometry, calculus)} &
\textcolor{mapsaccent}{Stepwise symbolic reasoning and scratchpad planning; optionally assisted by computational tools (e.g., Wolfram Alpha)} &
\textcolor{mapsaccent}{Exact match to final solution} \\[3pt]

\textcolor{mapsaccent}{ASB}~\citep{zhang2024agent} & \textcolor{mapsaccent}{Security and safety stress testing} &
\textcolor{mapsaccent}{Policy adherence, safe tool invocation, and memory integrity under adversarial and injected prompts} &
\textcolor{mapsaccent}{Attack success rate / refusal rate} \\
\bottomrule
\end{tabular}
}
\caption{\textcolor{mapsaccent}{Summary of benchmarks included in \name. Together, they span diverse domains and evaluation goals, enabling comprehensive multilingual analysis of agentic performance and security.}}
\label{tab:datasets}
\end{table*}

\textbf{Human Expert Verification.} To ensure this pipeline's reliability, we conducted human verification on a representative subset of translations. 
Evaluation setup and results are in Subsection~\ref{tran_imp}.

\subsection{Dataset Selection and Composition}

\textbf{Dataset Selection.} To enable robust multilingual evaluation of agentic capabilities, we construct \name benchmark suite from established agentic AI benchmarks.
These were selected based on four criteria: (1) strong research community adoption, including prior use in agentic evaluation; (2) clearly defined, closed-form answers for controlled evaluation; (3) sufficient difficulty to challenge open-source agents; and (4) practical solvability, ensuring that multilingual degradation can be measured. 
\textcolor{mapsaccent}{Accordingly, we selected four benchmarks spanning real-world reasoning (GAIA \citet{mialon2023gaia}), software engineering (SWE-Bench~\citet{jimenez2023swe}), mathematical problem solving (MATH \citet{hendrycks2021measuring}), and security assessment (Agent Security Benchmark, ASB \citet{zhang2024agent}).
Each dataset represents a distinct aspect of agentic behavior-planning, tool use, reasoning, and security. 
Table \ref{tab:datasets} summarizes their task types, evaluated capabilities, and metrics, with full dataset details provided in the Appendix.} 

\textbf{Data Composition}. 

\textit{Translated Languages.} We selected eleven typologically and geographically diverse languages:
German, Spanish, Portuguese (Brazil), Japanese, Russian, Chinese, Italian, Arabic, Hebrew, Korean, and Hindi, covering a broad range of scripts, linguistic structures, and regional user populations. Details on the specific dialects translated for each language are provided in the Appendix.

\textit{Data Handling.} To preserve the integrity of the original benchmarks, we applied only minimal targeted modifications. 
Translations were appended without altering original metadata (e.g., task type, difficulty level, or tool availability) and domain-specific syntax, such as equations in MATH, code in SWE-Bench, and adversarial prompts in ASB, was preserved exactly.
For MATH and SWE-Bench, which were not originally agentic, we retained only the most challenging tasks by difficulty level.
This follows common practice in prior work for aligning non-agentic datasets to agentic evaluation \citep{wu2023mathchat}, ensuring meaningful agent evaluation.

\textit{Data Volume.} To balance performance and security evaluation, \name comprises $805$ tasks: $405$ from performance-oriented datasets (GAIA, SWE-Bench, MATH) and $400$ from the ASB. 
Specifically, we include $165$ GAIA tasks (full validation set), $140$ high-difficulty MATH problems ($20$ per topic across $7$ topics), and $100$ medium-to-hard SWE-Bench tasks, alongside all $400$ ASB prompts.
Each task was translated into $11$ languages and combined with the original English version, yielding $9{,}660$ multilingual instances across $12$ languages.

To validate the benchmark’s utility and examine the \emph{Multilingual Effects}, we applied a leading agent to each dataset (see Section~\ref{evaluation} for details).


\begin{figure*}[t]
\includegraphics[width=1.0\textwidth]{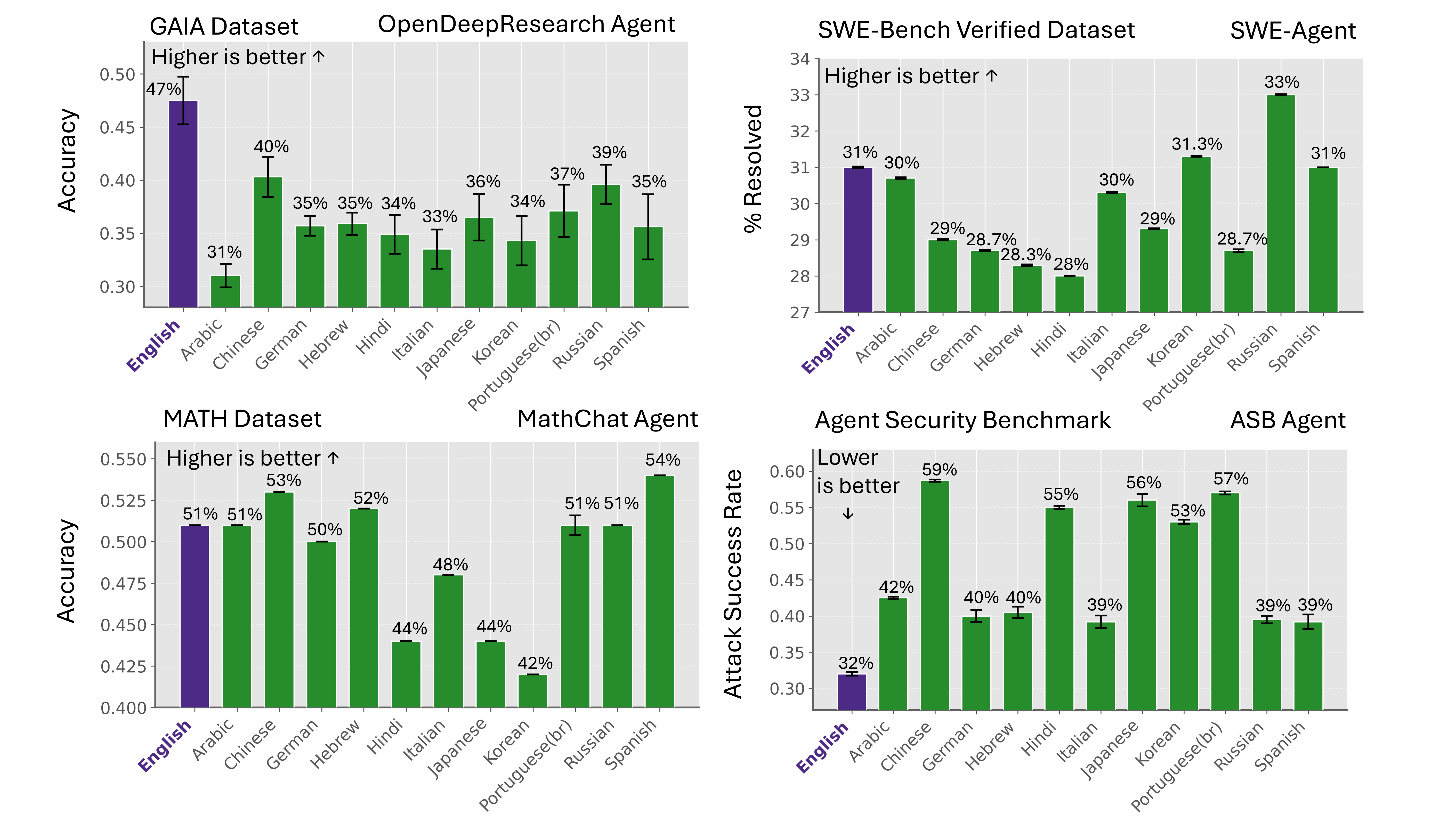}
\caption{Performance of open-source agents across languages on four agentic benchmarks: GAIA, SWE-Bench, MATH, and ASB. Each bar represents the agent’s accuracy (or attack success rate in ASB) for a given language, with English shown as the baseline. Error bars indicate std across three runs. Performance differences reflect each agent's degradation or resilience in multilingual settings.}
\label{fig:quant_results}
\end{figure*}

\textcolor{mapsaccent}{
\textbf{Design Rationale: Fixed Environments with Multilingual Inputs.}
\name models realistic deployment settings where users interact in non-English languages while the underlying environment (e.g., documentation, code, or tools) remains primarily English. Translating only the instruction side ensures:
(i) \textit{System compatibility:} alignment with current deployments interfacing with English-centric tools;
(ii) \textit{Cross-language comparability:} all languages face identical evidence and gold answers, allowing performance and security gaps to be attributed purely to language understanding rather than environmental differences. 
Agents may use English internally, avoiding artificial constraints.
Further discussion appears in Section \ref{lim}.}

\subsection{Translation Implementation and Verification}\label{tran_imp}
 \textbf{Translation Implementation Details.} 
For neural machine translation, we used Google Translate \cite{googletranslateapi}; for LLM translation, Cohere's Command A \cite{cohere2025command} with a designated system prompt. Structured expressions (e.g., code, URLs, equations) were identified via special tokens provided by the datasets \cite{zhang2024agent, hendrycks2021measuring}, and masked accordingly in the translation process.

\begin{figure*}[t]
\includegraphics[width=1.0\textwidth]{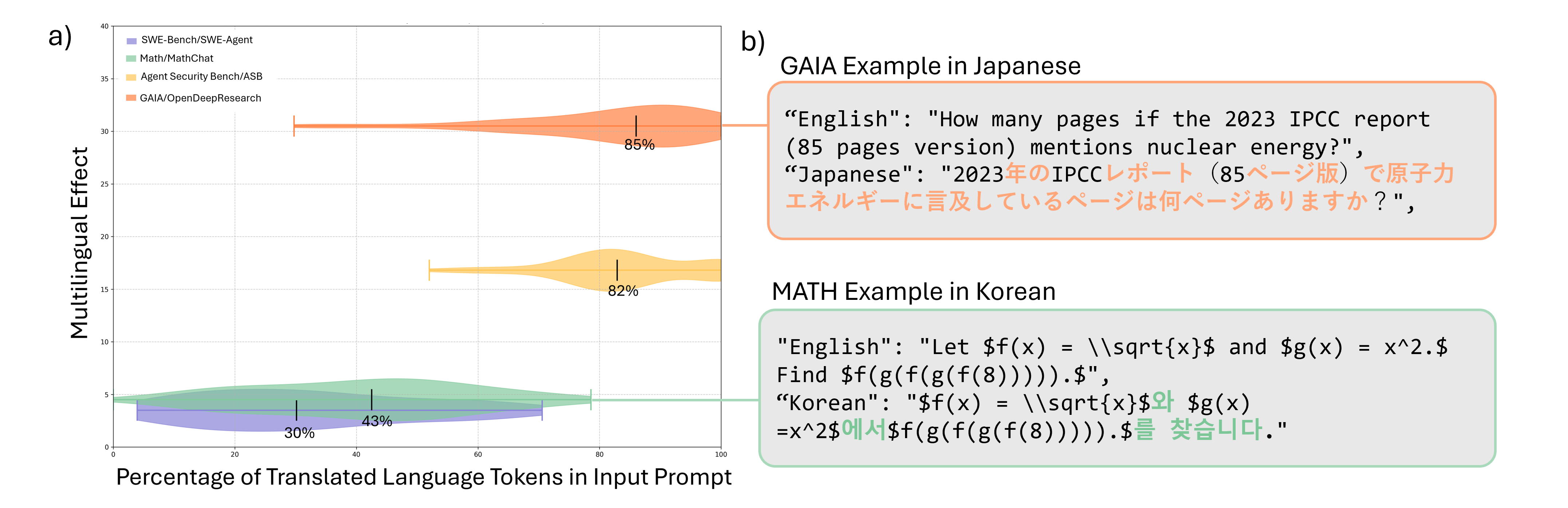}
\caption{a) Multilingual Effect as a function of the proportion of translated language tokens in input prompts. Each point represents a benchmark-agent pair, with the Multilingual Effect computed as the average relative degradation in performance or security across non-English languages. The trend suggests a correlation between input translation extent and multilingual vulnerability.
\textbf{b)} Two snippets exemplify a low‑translation prompt (MATH) and a high‑translation prompt (GAIA), clarifying the x‑axis percentages in panel (a) and showing how the proportion of natural‑language tokens, rather than task difficulty alone, drives the observed Multilingual Effect.}
\label{fig:additional_quan_results}
\end{figure*}

\textbf{Quality Assurance.}
Translation quality was verified by native-speaking annotators fluent in both English and the target language.
Each sample was rated on a 1–5 Likert scale across three established criteria: \textit{adequacy} (semantic fidelity), \textit{fluency} (grammatical and stylistic naturalness), and \textit{formatting accuracy} (preservation of LaTeX, code, etc.) \cite{freitag2021experts}.
A fourth metric, \textit{answerability}, measured whether the translation preserved intent well enough for the annotator to confidently answer the question as if it were in English.
Annotator instructions provided in the Appendix.
To validate verifiers' reliability, we embedded \textit{honeypot} samples with intentional errors; annotators reliably flagged these cases, confirming quality control.

Evaluation results confirm high translation quality, with an answerability rate of $94.2$\%, and mean scores of $4.43$ (adequacy), $4.59$ (fluency), and $4.75$ (formatting) on the 1–5 Likert scale, demonstrating strong semantic fidelity, linguistic naturalness, and structural integrity.
A total of 2,200 samples (25\% of the benchmark) were manually verified, covering all languages and task types. 
At this verification scale, statistical analysis shows that even under the most conservative assumptions, the 95\% confidence interval for the true corpus-level answerability rate is \([92.0,\,96.4]\%\).
This suggests that evaluating a quarter of the dataset provides a reasonably precise estimate of overall translation quality, with limited sampling uncertainty.
Full per-language results and analysis are included in the Appendix. To support high-precision use cases, we also release a “verified” subset, consisting of approximately $190$ translations per language that passed native expert review. 

\textcolor{mapsaccent}{
\textbf{Answerability as key metric.} While adequacy and fluency confirm general translation quality, they do not ensure reliable task execution.
Our correlation analysis shows that \textit{answerability}, the extent to which a translation remains solvable, most closely aligns with correctness, whereas adequacy and fluency are not predictive. 
Although this relationship is not absolute since performance naturally also depends on task complexity, answerability remains the most reliable indicator of translation fidelity across languages (see Appendix for details).}

\section{Experiments}\label{evaluation}


\subsection{Experimental Settings}

\textbf{Agent assignment per benchmark.} To demonstrate the utility of MAPS, we evaluate one strong, established open-source agent per benchmark, each representing the current state of practice in its respective domain. While a single unified agent would offer a more controlled comparison, such general-purpose agents do not yet exist at sufficient maturity: contemporary agentic systems are tightly specialized around particular toolchains, environments, and prompting strategies, and transferring them across domains typically results in severe performance collapse \citep{gioacchini2024agentquest,chang2024agentboard}. 
Importantly, our goal is not to compare agent architectures, but to measure the effect of language on agentic behavior within a fixed agent–environment pair. For each benchmark, the agent, tools, prompts, and LLM backbone and configuration are held constant, and only the input language is varied. This isolates multilinguality as the sole experimental factor, ensuring that any observed degradation reflects language-conditioned failure modes rather than architectural differences.

For GAIA, we employ the OpenDeepResearch (ODR) agent \citep{huggingface2024openresearch}, integrating retrieval, web browsing, and tool use for real-world reasoning.
For MATH, we use MathChat~\citep{wu2024mathchat}, a zero-shot multi-turn reasoning agent with Python execution and the Wolfram Alpha tool. 
For SWE-Bench, we applied SWE-agent~\citep{yang2024swe} for autonomous code reasoning through repository navigation, file editing, and test execution. 
For ASB, we built on the original benchmark infrastructure~\citep{zhang2024agent}, eliciting synthetic adversarial tool-use in various agentic scenarios (e.g., travel, legal, system admin).
ODR runs with \emph{GPT-o1}, MathChat with \emph{GPT-4}, SWE-Agent with \emph{GPT-4.1}, and ASB with \textcolor{mapsaccent}{\emph{GPT-5-mini}}. Full configuration details are in the Appendix.

\textbf{Experiment Protocol.} Each agent was evaluated three times per language, across all $12$ languages ($36$ runs per benchmark). We report mean and standard deviation over these runs in Fig.~\ref{fig:quant_results}.


\textbf{Metrics.} We adopt each benchmark’s original evaluation metrics for consistency with prior work:
accuracy for MATH and GAIA, test-case success rate for SWE-Bench, and attack success rate (ASR) for ASB.
To enable cross-benchmark comparison of multilingual degradation, we further define a unified scalar measure, which we call the Multilingual Effect. For a given benchmark–agent pair, this is computed as the difference between the English performance score and the mean performance score across the 11 non-English languages. For ASB, where higher scores indicate worse security, the sign is inverted so that larger values consistently indicate greater degradation.
Intuitively, the Multilingual Effect quantifies how much performance or safety is lost solely due to changing the interaction language. We use this measure throughout the analysis to study how linguistic load, task structure, and agent design modulate multilingual vulnerability.

\subsection{Results}
Fig. \ref{fig:quant_results} presents agent performance across all four benchmarks in English and the eleven target languages. 
In GAIA and ASB, we observe a clear Multilingual Effect: non-English languages consistently underperformed compared to English, with reductions of up to 16\% (GAIA) and a vulnerability increase of up to \textcolor{mapsaccent}{27\%} in ASB. 
Notably, SWE-Bench and MATH exhibit only minor variation, with scores clustered around the English baseline.

These results reveal important differences in how multilingual degradation manifests across task types. Although all tasks require complex reasoning, some are more constrained than others. 
For instance, SWE-Bench and MATH primarily involve structured code or mathematical expressions, thus placing less emphasis on natural language and more on formal syntax or notation, thereby reducing the Multilingual Effect.
In contrast, GAIA focuses on solving real-world tasks where understanding flexible natural language prompts is essential. Thus, the importance of the natural language problem statement is significantly higher. 
To explore this variation, we examine the proportion of localized, target-language text in each benchmark’s input (Fig.~\ref{fig:additional_quan_results}).



Fig. \ref{fig:additional_quan_results} examines the relationship between prompt composition and multilingual performance. 
Part (a) shows a correlation between the percentage of non-English tokens in the input and the average performance gap (relative to English).
Benchmarks with higher proportions of localized, target-language-oriented input, such as GAIA and ASB, exhibit greater degradation, whereas SWE-Bench, with predominantly English input (e.g., code), shows higher preservation.
Part (b) contrasts a minimally translated MATH sample example with a nearly fully translated GAIA query


\begin{table}[t]
\centering
\scriptsize
\setlength{\tabcolsep}{3.5pt} 
\begin{tabular}{l|c|ccc}
\toprule
 & \textbf{English} & \multicolumn{3}{c}{\textbf{Mean (Non-English)}} \\
\textbf{} & & Orig. & \textcolor{mapsaccent}{Determin. trans.} & \textcolor{mapsaccent}{Agent-dec. trans.} \\
\midrule
GAIA (ACC) $\uparrow$ & 47.4\% & 35.8\% & \textcolor{mapsaccent}{42\%} & 40.9\% \\
ASB (ASR) $\downarrow$ & \textcolor{mapsaccent}{32.7\%} & \textcolor{mapsaccent}{47.3\%} & \textcolor{mapsaccent}{40.4\%}& \textcolor{mapsaccent}{43.3\%} \\
\bottomrule
\end{tabular}
\caption{\textcolor{mapsaccent}{Comparison of deterministic and agent-decision translation on GAIA and ASB. We evaluate two ways to make the agent \emph{aware} of multilingual context: (1) Deterministic translation, where inputs are pre-translated to English, and (2) Agent-decision translation, where the agent decides if and how to translate. Both recover part of the lost performance and robustness, though a clear gap to the English baseline remains.}}
\label{tab:single-col-results}
\end{table}

\textcolor{mapsaccent}{\textbf{Translation-based ablations.}
To disentangle translation effects from genuine multilingual degradation, we evaluate two strategies (Table~\ref{tab:single-col-results}). In \emph{deterministic translation}, all non-English prompts are translated to English before inference, isolating translation quality effects. In \emph{agent-decision translation}, the agent decides if and how to translate, reflecting realistic multilingual deployment. Deterministic translation recovers roughly 5\% of the lost performance - consistent with our 5.8\% human-verified translation error, while the multilingual gap remains higher (GAIA 12\%, ASB 14\%), indicating that \emph{translation bias} alone cannot explain the degradation. 
The \emph{agent-decision} variant yields comparable gains, showing partial self-correction through contextual awareness. Yet, both stay below the English baseline, pointing to deeper agent-level reasoning limits across languages.}

\begin{table}[t]
\centering
\scriptsize
\renewcommand{\arraystretch}{1.05}
\setlength{\tabcolsep}{3pt}
\resizebox{\columnwidth}{!}{
\textcolor{mapsaccent}{
\begin{tabular}{llcc}
\toprule
\textbf{Setting} & \textbf{Benchmark(s)} & \textbf{Configuration} & \makecell{\textbf{Multilingual}\\\textbf{Degradation (\%)}} \\
\midrule
\multicolumn{4}{c}{\textbf{Cross-Benchmark Check – Same Agent, Different Benchmarks}} \\
\midrule
ODR & \textit{MATH} & GPT-o1 & –0.1\% \\
MATHChat & \textit{MATH} & GPT-4 & –0.1\% \\
ODR & \textit{GAIA} & GPT-o1 & –12\% \\
\midrule
\multicolumn{4}{c}{\textbf{Multiple LLM Check – Same Agent, Different Base LLMs}} \\
\midrule
ODR & \textit{GAIA} & GPT-o1, GPT-4o  &  –12\%,–10\% \\
ASB Agent & \textit{ASB} & GPT-5-mini, Qwen-2  &   –14.6\%,–8.8\% \\
\bottomrule
\end{tabular}
}}
\caption{\textcolor{mapsaccent}{Cross-benchmark and cross-model analysis. Results indicate relative degradation compared to the English baseline. Multilingual performance remains stable in language-light tasks (\textit{MATH}), but consistently degrades in language-heavy benchmarks (\textit{GAIA}, \textit{ASB}), independent of the underlying agent or base model.}}
\label{tab:cross-generalization}
\end{table}

\textcolor{mapsaccent}{\textbf{Cross-Benchmark Evaluation.}
We examined whether multilingual degradation stems from agent designs or broader task characteristics. Such analyses are constrained by the \emph{benchmark-overfitting problem}, where agents are tightly coupled to domain-specific tools and prompts, limiting cross-task generalization. We identified one viable case: using \emph{OpenDeepResearch} (ODR), our GAIA agent, on the language-light \textit{MATH} benchmark. ODR matched MathChat’s accuracy with no multilingual drop, yet degraded by 12\% on GAIA, indicating that degradation arises primarily from linguistic load rather than agent architecture (see Table~\ref{tab:cross-generalization}).}

\textcolor{mapsaccent}{\textbf{Cross-Model Evaluation.}
We next tested whether the multilingual degradation effect is specific to a particular LLM. Replacing \emph{GPT-o1} with \emph{GPT-4o} for ODR and \emph{GPT-5-mini} with \emph{Qwen-2} for ASB yields the same pattern: performance drops in GAIA and elevated ASR in ASB. 
These consistent trends suggest that multilingual degradation reflects a general property of language-conditioned reasoning, rather than a model-specific bias.}

\textbf{Unified-backbone check.} A remaining concern is that multilingual differences across benchmarks might be caused by assigning different base LLMs to different agents (e.g., GPT-o1 for GAIA, GPT-4.1 for SWE-Agent, GPT-5-mini for ASB). While our primary setup preserves each agent’s native configuration to avoid unrepresentative failures (Section 4.1), we perform an additional experiment in which all agents are re-evaluated using the same backbone LLM, GPT-5-mini, while keeping all other components (agent logic, prompts, tools, and environments) fixed. Table \ref{tab:unified_backbone} shows a re-evaluation of all benchmarks under this unified setting, which yields the same qualitative multilingual trends: GAIA and ASB continue to exhibit substantial multilingual degradation, whereas MATH and SWE-Bench remain comparatively stable across languages. Although absolute performance improves due to the stronger base model, the relative multilingual gaps persist, confirming that cross-benchmark differences are driven by task structure rather than model assignment. Full per-language results are provided in the Appendix.

\begin{table}[t]
\centering
\footnotesize
\setlength{\tabcolsep}{4pt}
\begin{tabular}{lcc}
\hline
\textbf{Benchmark} & \textbf{Original} & \textbf{Unified} \\
\hline
GAIA (Acc $\uparrow$) & $-12\%$ & $-10\%$ \\
ASB (ASR $\downarrow$) & $+14.6\%$ & $+14.6\%$ \\
MATH (Acc $\uparrow$) & $-0.1\%$ & $-0.9\%$ \\
SWE-Bench (Acc $\uparrow$) & $-1.3\%$ & $-0.3\%$ \\
\hline
\end{tabular}
\caption{Multilingual Effect (relative degradation vs.\ English) under the original agent setup and under a unified backbone (GPT-5-mini). Language-heavy benchmarks (GAIA, ASB) degrade substantially, while language-light benchmarks (MATH, SWE-Bench) remain stable.}
\label{tab:unified_backbone}
\end{table}

\begin{figure}[h]
\includegraphics[width=0.49\textwidth]{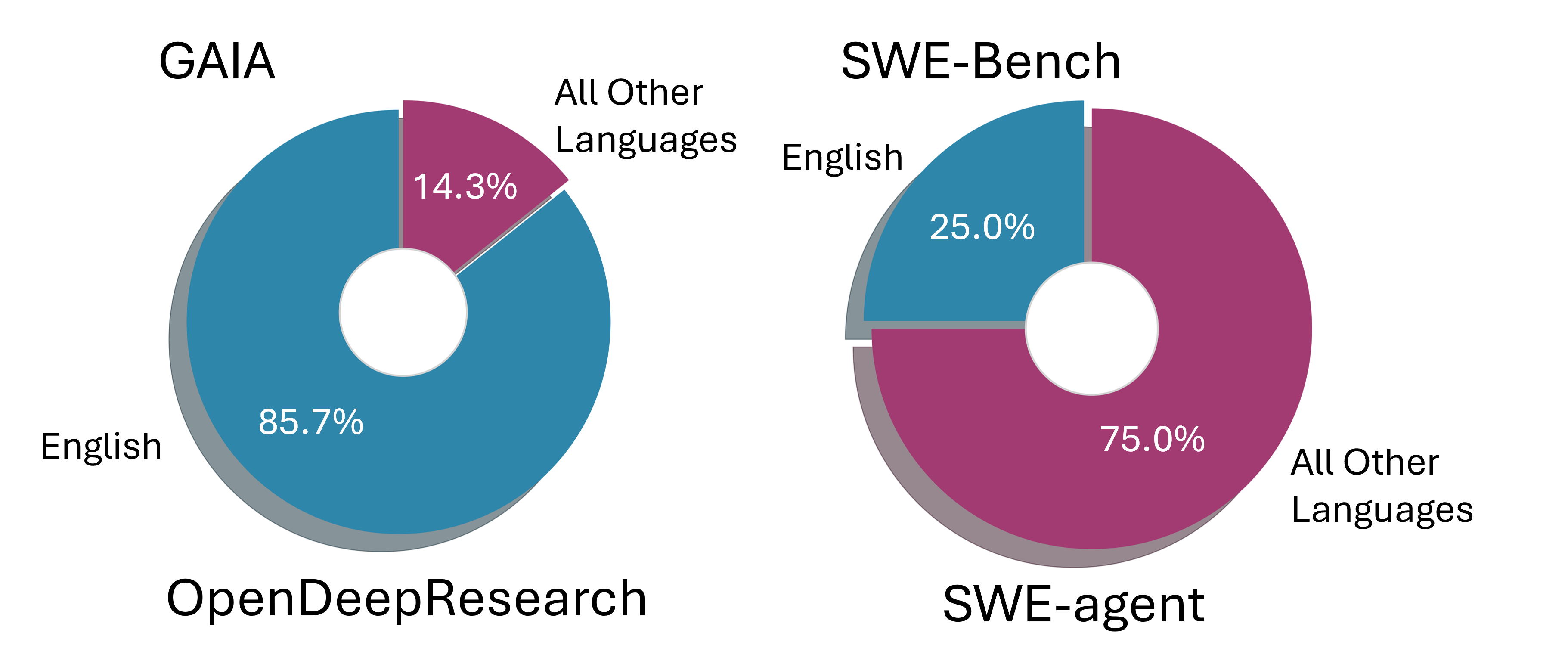}
\caption{Cross-lingual planning consistency for OpenDeepResearch (ODR) and SWE-Agent. For each agent, we locate tasks solved in English but failed in all other languages, then measure the semantic similarity between the English instruction and each language’s initial planning step using multilingual embeddings. ODR exhibits a strong cross-lingual gap: in 85.7\% of cases, its English planning step is more faithful to the instruction than in other languages. In contrast, SWE-Agent is more robust, with English leading in only 25\% of cases.}
\label{fig:pie_chart}
\end{figure}

\textbf{Cross-Lingual Failure Analysis}
To identify the \textit{root causes} of multilingual degradation, we conduct a deeper analysis of agents’ \textit{initial planning behavior} across languages—the first stage where an instruction is decomposed into a sequence of actions.
Using multilingual embeddings \cite{openai_embeddings_v3}, we measure the cosine similarity between each language’s plan and its English counterpart, quantifying how closely non-English reasoning aligns with the original task intent.
We focus on tasks solved in English but failed in all other languages, comparing two contrasting agents: \emph{ODR} (GAIA) and \emph{SWE-Agent} (SWE-Bench)—which show the strongest and weakest Multilingual Effects, respectively (Fig.~\ref{fig:additional_quan_results}).

As shown in Fig.~\ref{fig:pie_chart}, ODR displays a clear English bias: its planning steps align more closely with task instructions in 85\% of cases. In contrast, SWE-Agent shows minimal cross-lingual variation, indicating stable reasoning across languages.
These results suggest that when an agent’s initial planning is language-invariant, multilingual degradation diminishes. A qualitative example of such an error is shown in the appendix.

\section{Discussion}\label{dis}

\subsection{Guidelines for Multilingual Agent Deployment and Risk Assessment}

\textbf{Guidelines for Language-Aware Deployment.} Before deploying an AI agent in a multilingual setting, analyze the linguistic composition of its expected input, particularly the balance between structured elements (e.g., code, formal queries) and localized natural language.
Inputs with a high proportion of natural language tend to increase the risk of performance and safety degradation.
We therefore recommend that, in such a case, developers should conduct a Multilingual Benchmark Assessment using a diverse, language-sensitive evaluation suite, such as \name, for AI agents operating across languages. This helps reveal vulnerabilities and promotes reliable global real-world behavior.


\textbf{Multilingual Inputs Amplify Agentic Security Vulnerabilities.} 
Our ASB evaluation revealed that multilingual adversarial inputs can bypass agent safety mechanisms with minimal sophistication. 
Direct translations of English jailbreak prompts, without any obfuscation, were sufficient to induce policy-violating behavior across languages, \textcolor{mapsaccent}{exposing latent vulnerabilities in agent pipelines. Even advanced models such as \textit{GPT-5-mini} proved more resilient in English than in translation.
We thus recommend treating multilingual robustness as a \emph{security-critical dimension}, and incorporating translated attacks as a standard component of agentic red-teaming risk assessments.}


\textcolor{mapsaccent}{
\subsection{Agentic Robustness: Beyond LLM Multilingual Evaluation.}
Our results show that multilingual robustness cannot be inferred from LLM-only benchmarks, since agentic systems introduce distinct failure modes absent from text generation.
Even when an underlying LLM remains linguistically competent, agents can still fail when language misinterpretations propagate into \emph{operational} errors such as planning drift or tool misuse (Fig.~\ref{fig:pie_chart}), reducing end-to-end reliability.
Accordingly, multilingual robustness must be evaluated directly at the \emph{agent level}, where performance and safety reflect language-conditioned decision making rather than text quality alone.
}

\section{Conclusions}\label{conc}
We introduce \name, the first multilingual benchmark suite for agentic AI systems, addressing a critical gap in assessing language-specific performance and security limitations. By extending four widely used agentic benchmarks—GAIA, SWE-Bench, MATH, and ASB—into eleven diverse languages, our suite enables the analysis of agent behavior under multilingual conditions.
Constructed through a hybrid translation pipeline and human verification, \name ensures high linguistic and structural consistency. Experiments reveal critical gaps when agents operate in non-English settings, particularly in tasks involving natural language reasoning and safety-critical behavior.
These findings underscore the importance of language-aware evaluation and multilingual adaptation, especially for real-world agentic deployments. We view \name as a practical foundation for building more inclusive, resilient, and globally reliable agentic AI systems, and invite the community to extend it.


\section{Limitations}\label{lim}
 
\name introduces the first multilingual suite for evaluating agentic AI systems, spanning four domains and eleven languages.
While this represents a strong foundation, expanding into additional domains—such as healthcare, or legal reasoning, and into extremely low-resource languages (e.g., Amharic, Uyghur) would broaden its applicability and strengthen its diagnostic power.

\textcolor{mapsaccent}{
A more specific limitation arises from our decision to vary only the input language, while keeping the agent environments fixed in English.
This design reflects the dominant real-world regime where users interact in multiple languages, but the underlying resources—documentation, codebases, schemas, or web content—remain primarily English.
While this setup simplifies comparison and isolates linguistic understanding as the primary variable, it does not capture the full complexity of localized ecosystems.
Fully non-English environments could introduce additional challenges stemming from tool compatibility, dataset structure, or culturally specific context.
Nevertheless, our fixed-environment design offers clear advantages: it ensures cross-language comparability, maintains consistent evidence and outputs, and provides a clean diagnostic for language-conditioned reasoning.
We therefore view MAPS as an essential first step toward localized multilingual evaluation—an upper bound on agent robustness that establishes a stable foundation for future extensions.}

\begingroup
\small
\bibliography{references}
\endgroup


\appendix

\section{Appendix}
\label{sec:appendix}

\begin{abstract}
This supplementary material provides additional details supporting our MAPS benchmark, including extended benchmark comparisons, experimental protocols, and multilingual evaluation results. It is organized as follows:
\begin{itemize}
    \item \textbf{Reproducibility Resources} – links to the full datasets and codebase, to enable open use and distribution for research and development purposes.
  \item \textbf{Agentic AI Benchmark Comparison} – an expanded analysis of existing benchmarks across dimensions such as evaluation objective, task type, answer format, and multilingual support.
  \item \textbf{Hybrid Translation Pipeline} – additional information on our hybrid translation methodology, including domain-specific prompt design and translation configuration experiments.
  \item \textbf{Manual Evaluation of Translation Quality} – annotation guidelines, evaluation examples, and detailed human verification scores per dataset and language.
  \item \textbf{Experimental Settings} – full setup specification, agent configurations, and language adaptation process used to isolate Multilingual Effects across tasks.
  \item \textbf{Extended Results} – detailed multilingual performance tables for GAIA, MATH, SWE-Bench, and ASB across runs.
\end{itemize}
\end{abstract}

\subsection{Reproducibility Resources}
We release the full \textbf{MAPS} dataset, which includes translations of four agentic AI benchmarks (GAIA, SWE-Bench, Math, and Agent Security Benchmark) across 12 languages. A verified subset of MAPS, manually reviewed by native speakers, is also included for higher-quality evaluations.
Links to both the full dataset and verified subset can be found at: \url{https://huggingface.co/datasets/Fujitsu-FRE/MAPS}.
We also provide the code for our hybrid translation pipeline, including prompt design and format-preserving techniques, available at: \url{https://github.com/omerhof-fujitsu/hybrid_translation_demo/}.

\begin{table*}[t]
\centering
\renewcommand{\arraystretch}{1.2}
\begin{adjustbox}{width=\textwidth}
\begin{tabular}{|l|c|c|l|c|c|c|c|}
\hline
\textbf{Paper} & \textbf{Evaluation Objective} & \textbf{Task Scope} & \textbf{Domain Coverage} & \textbf{\# of tasks} & \textbf{Answer Format} & \textbf{GT Available} & \textbf{Multilingual} \\
\hline
\textbf{\textit{GAIA \citep{mialon2023gaia}}} & Performance & Full-agentic & General, multimodal & 460 & Close & \cmark & \xmark \\
\textbf{\textit{SWE Bench \citep{jimenez2023swe}}} & Performance & Full-agentic & Software engineering & 3000 & Close & \cmark & \xmark \\
\textbf{\textit{MATH \citep{hendrycks2021measuring}}} & Performance & Full-agentic & Math solver & 12,500 & Close & \cmark & \xmark \\
\textit{$\tau$-bench \cite{yao2406tau}} & Performance & Semi-agentic & Airline, retail & 165 & Close & \cmark & \xmark \\
\textit{AgentBoard \citep{chang2024agentboard}} & Performance & Semi-agentic & General – 9 varied tasks & 134 & Close & \cmark & \xmark \\
\textit{App World \citep{trivedi2024appworld}} & Performance & Semi-agentic & Apps, code execution, APIs & 750 & Close & Partial & \xmark \\
\textit{TheAgentCompany \citep{xu2024theagentcompany}} & Performance & Semi-agentic & SW Dev, HR, Admin, DS & 175 & Close & \cmark & \xmark \\
\textit{ARA \citep{kinniment2023evaluating}} & Performance, Security & Semi-agentic & SW Dev, phishing, web searches & 12 & Open & \xmark & \xmark \\
\textit{TravelPlanner \citep{xie2024travelplanner}} & Performance & Full-agentic & Travel planner, tourist & 1225 & Open & \xmark & \xmark \\
\hline
\end{tabular}
\end{adjustbox}
\caption{Comparison of existing agentic AI benchmarks along evaluation objective, scope, domain coverage, and multilingual availability.}
\label{benchmark_comparison}
\end{table*}

\begin{table*}[t]
\centering
\renewcommand{\arraystretch}{1.2}
\resizebox{\textwidth}{!}{%
\begin{tabular}{|l|p{1.5cm}|c|l|c|c|c|c|}
\hline
\textbf{Paper} & \textbf{Evaluation\newline Objective} & \textbf{Task Scope} & \textbf{Domain Coverage} & \textbf{\# of tasks} & \textbf{Answer Format} & \textbf{GT Available} & \textbf{Multilingual} \\
\hline
\textbf{\textit{Agent Security Bench \citep{zhang2024agent}}} & Security & Semi-agentic & Offensive cyber operations, unsafe tool & 90,000 & Close & \cmark & \xmark \\
\textit{AgentHarm \citep{andriushchenko2410agentharm}} & Security & Full-agentic & Fraud, cybercrime, harassment & 110 & Open & \xmark & \xmark \\
\textit{AgentDojo \citep{debenedetti2024agentdojo}} & Security & Semi-agentic & Prompt injection, phishing, data leakage & 97 & Both & Partial & \xmark \\
\textit{InjecAgent \citep{zhan2024injecagent}} & Security & Semi-agentic & Indirect prompt injection, data leakage & 1,000 & Close & \cmark & \xmark \\
\textit{AgentPoison \citep{chen2024agentpoison}} & Security & Full-agentic & Backdoor attacks, data poisoning & \textasciitilde20 & Open & \xmark & \xmark \\
\textit{BIPIA \citep{yi2023benchmarking}} & Security & Semi-agentic & Indirect prompt injection, data leakage & 5 & Close & \cmark & \xmark \\
\textit{Evil Geniuses \citep{tian2023evil}} & Security & Full-agentic & Prompt injection, data leakage & 156 & Open & \xmark & \xmark \\
\hline
\end{tabular}%
}
\caption{Comparison of agentic security benchmarks by scope, threat type, answer format, and multilingual coverage.}
\label{security_benchmark_comparison}
\end{table*}

\subsection{Agentic AI Benchmark Comparison}
To contextualize \name within the growing landscape of agentic evaluation, we provide a detailed comparison of recent benchmarks across core design dimensions. 
While the main paper (Section 2.1) outlines a conceptual taxonomy—covering evaluation objectives, agent autonomy levels, and domain specialization—this supplementary section complements it with a structured, side-by-side table. 
Our comparison highlights key properties such as the ground truth availability, domain coverage, and multilingual support. 
Notably, \name remains the only benchmark to offer multilingual assessment across multiple agentic tasks, spanning both performance and safety. 
This contrast underscores an existing gap in multilingual agent evaluation and illustrates how \name fills it.

The columns in Tables \ref{benchmark_comparison} and \ref{security_benchmark_comparison} summarize key properties of each benchmark as follows. \textbf{Evaluation Objective} specifies what the benchmark is primarily designed to measure. Performance benchmarks evaluate the agent's accuracy to fulfill the given tasks, whereas Security benchmarks probe robustness to adversarial inputs. \textbf{Task Scope} describes the degree of freedom given to the agent. Fully agentic benchmarks specify only the task and success criterion, allowing arbitrary agent architectures and strategies, whereas semi-agentic benchmarks provide a partially fixed agentic scaffold (e.g., code templates or tool interfaces) and mainly vary the underlying LLM or prompting. \textbf{Domain Coverage} indicates the application domains spanned by the benchmark (e.g., software engineering, mathematics, web navigation, security). \textbf{ of tasks} reports the number of evaluation instances included in the benchmark. \textbf{Answer Format} specifies how solutions are expressed and evaluated. Closed formats require producing a result in a predefined, structured form with a definitive correct answer, whereas open formats allow free-form outputs for which multiple solutions may be valid. \textbf{Ground Truth Available} indicates whether each task has an explicit reference solution. \textbf{Multilingual} denotes whether the benchmark natively supports evaluation in multiple languages, as opposed to being designed only for English inputs.

Table \ref{benchmark_comparison} surveys general-purpose agentic benchmarks, revealing that most focus exclusively on performance, operate within semi-agentic environments, and lack multilingual coverage. 
Full-agentic settings—those requiring autonomous planning and execution without scaffolding—are limited to a few benchmarks such as \textit{GAIA}, \textit{SWE-Bench}, and \textit{MATH}, each addressing distinct domains: multimodal reasoning, software engineering, and mathematical problem solving, respectively. 
Table \ref{security_benchmark_comparison} focuses on security-oriented agentic benchmarks. 
While these benchmarks advance robustness evaluation, most adopt semi-agentic setups and open-ended answer formats, and often withhold ground truth to prevent overfitting.

Based on this analysis, we selected four complementary benchmarks—\textit{GAIA}, \textit{SWE-Bench}, \textit{MATH}, and \textit{Agent Security Bench (ASB)}—that collectively span performance and security objectives, cover diverse domains including real-world reasoning, code, math, and cybersecurity, and offer closed-form outputs with accessible ground truth. Notably, three of the four operate in fully agentic settings, aligning with our emphasis on end-to-end autonomous decision-making. 
This configuration enables reliable, multilingual evaluation across both agent capabilities and vulnerabilities, providing a robust foundation for \name. 

\textcolor{mapsaccent}{Below, we provide additional details regarding each benchmark:}

\textcolor{mapsaccent}{\textit{GAIA.} GAIA \citep{mialon2023gaia} is a benchmark designed to evaluate agents' performance on real-world assistant tasks.
It includes curated questions that require multi-step reasoning and autonomous use of tools such as web browsers, code interpreters, or document analyzers. 
Each question has a single correct answer, and responses are evaluated by an exact match to a reference output.}

\textcolor{mapsaccent}{\textit{SWE-Bench.} SWE-Bench~\citep{jimenez2023swe} is a software engineering benchmark constructed from real GitHub issues and associated pull requests across popular Python repositories. 
Each task presents a bug report and a codebase snapshot, and requires the agent to evaluate whether a proposed patch correctly resolves the issue. 
We adopt the \emph{verified} subset \cite{openai2024swebench}, in which agents are tasked with validating a patch rather than generating one.} 

\textcolor{mapsaccent}{\textit{MATH.} The MATH dataset \citep{hendrycks2021measuring} includes high-school level mathematical problems across seven topics, including algebra, geometry, and calculus. 
Tasks are structured to require symbolic manipulation and multi-step reasoning. 
Agent responses are evaluated by exact match against a reference solution.}

\textcolor{mapsaccent}{\textit{Agent Security Benchmark (ASB).} ASB benchmark \citep{zhang2024agent} provides a structured evaluation of agent robustness against adversarial threats, including prompt injections, memory poisoning, and tool misuse. Agents interact with injected prompts or environments, and evaluation is based on whether safety policies are violated, measured by attack success rate and refusal rate.
}

\subsection{Hybrid Translation Pipeline Implementation Additional Information}

\begin{figure*}[t]
    \centering
    \includegraphics[width=\textwidth]{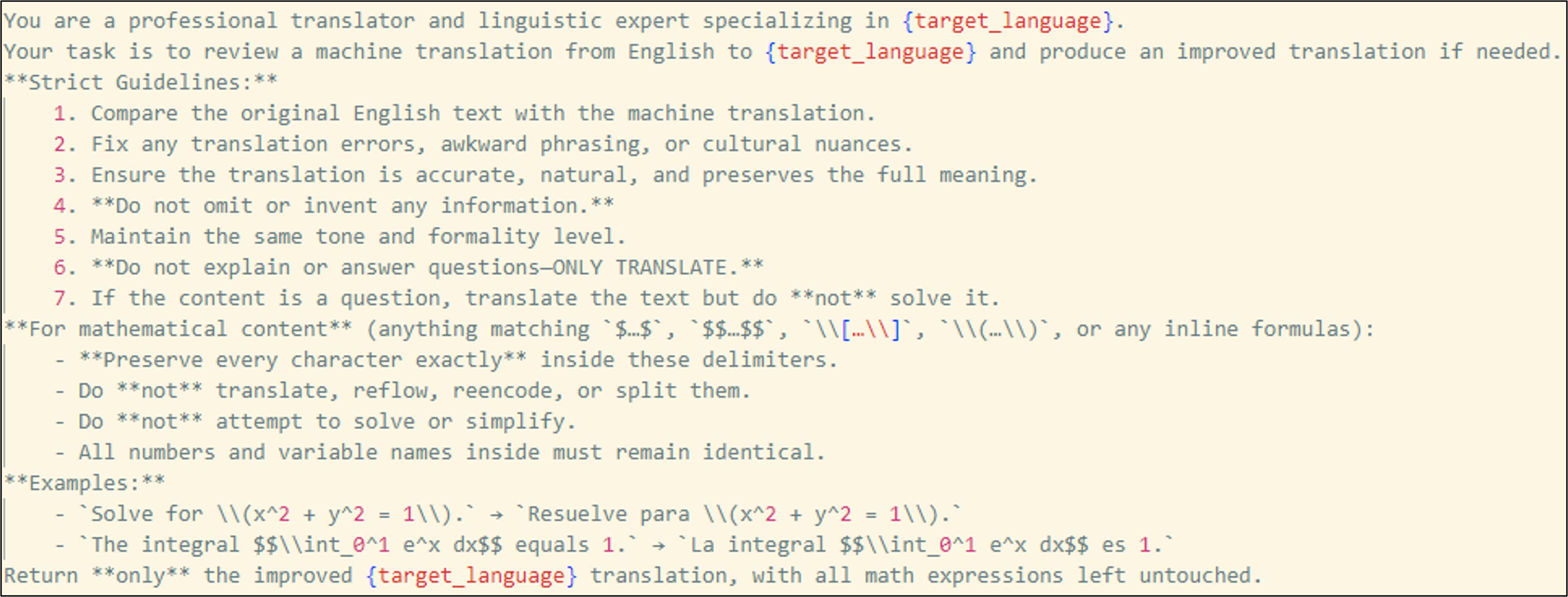}
    \caption{System prompt used for math translation refinement. This prompt is part of the hybrid translation pipeline and instructs the model to improve machine-translated math questions while preserving all LaTeX expressions and symbolic structure. As part of our hybrid translation pipeline, we designed domain-specific prompts for each benchmark to ensure appropriate evaluation and structural fidelity.}

    \label{fig:math_machine_translation_verification_prompt}
\end{figure*}

Reliable multilingual evaluation of agentic AI systems hinges on translating task instructions with both semantic and structural cross-language fidelity. 
Neural MT excels at preserving format and structure, but struggles in low-resource or specialized domains \citep{koehn2017six,aharoni2019massively}. 
Translation via instructed LLMs offers broader high-level capabilities at the cost of occasional hallucinations and semantic drift \citep{hendy2023good,yan2024gpt}. To balance these trade-offs, hybrid pipelines were suggested by \citet{ki2024guiding,mohammed2025google}, combining format-preserving MT with LLM-based refinement. 
For \name, we extend \citet{ki2024guiding}: First, \citet{ki2024guiding} was not designed with our benchmarks in mind, thus significant per-benchmark prompting had to be done. 
Second, we added automated quality checks, fallbacks, and expert verification to ensure the cross-language fidelity needed for agentic benchmarks (Fig.~\ref{fig:translation_pipe}). 

Formally, let us express our translation pipeline as a function $T: \mathcal{S} \times \mathcal{L} \rightarrow \mathcal{T}$, where $s \in \mathcal{S}$ is a task-instruction instance in source-language (English), $L_t \in \mathcal{L}$ is the target language, and $t \in \mathcal{T}$ is the resulting translated output. 
The pipeline begins with machine translation (MT) to establish a structural foundation: Denote $M(s, L_t)$, the MT function, implemented as a high-quality, off-the-shelf NMT system. Its output provides a structurally faithful baseline for subsequent steps. 

Next, we apply a verification step using an LLM to assess whether the translation adequately preserves the source meaning. This is modeled as a binary function $A(s, M(s, L_t), L_t) \rightarrow {0, 1}$, where the LLM compares the original and translated texts to detect major semantic errors or omissions. 

Based on verification outcomes, the pipeline follows one of two distinct paths.
 If $A = 0$ (indicating MT failure), the pipeline employs direct LLM translation: 
Denote $\Phi_{\text{direct}}(s, L_t)$ the output of an LLM prompted to directly translate $s$ to language $L_t$ (without using the MT output).
If $A = 1$ (indicating acceptable MT), an LLM enhances the translation while preserving its basic structure: Denote $\Phi_{\text{enhancement}}(s, M(s, L_t), L_t)$ as the output of an LLM, guided to improve the MT output while maintaining structural consistency.

To ensure semantic integrity, we apply a second binary check: $I(s, \Phi_{\text{enhancement}}) \rightarrow {0, 1}$
, targeting common LLM failure modes, such as hallucinations, omissions, misinterpretations (e.g., answering instead of translating), and semantic drift. 
If this verification fails, we revert to the original MT output (which passed the initial verification test).

These conditional steps form a robust decision framework: If MT is rejected, we use a direct LLM translation; if accepted but enhancement fails integrity verification, we fall back to the MT; Otherwise, we use the enhanced translation. 
Importantly, since both MT and LLM-based translation are robust technologies, we did not encounter any scenario in which both systems failed to produce a valid translation.
A formal expression of this pipeline is provided in the Appendix.

For the translation pipeline, we used Python 3.11 alongside a hybrid setup comprising Google’s Neural Machine Translation (NMT) API, Cohere Command-A, and GPT-4o. To support high-quality multilingual alignment, we crafted dedicated system prompts for each stage of the pipeline—translation verification, semantic enhancement, and structural integrity checks—tailored to the specific requirements of each domain. For example, GAIA tasks involve real-world decision-making, where preserving subtle intent and contextual detail across languages is critical. In contrast, MATH tasks require precise handling of symbolic notation and mathematical terminology, necessitating strict preservation of LaTeX expressions and domain-appropriate language. An example of a task-specific system prompt is provided in Fig.~\ref{fig:math_machine_translation_verification_prompt}. 

\begin{figure*}[t!]
    \centering
    \includegraphics[width=0.8\textwidth]{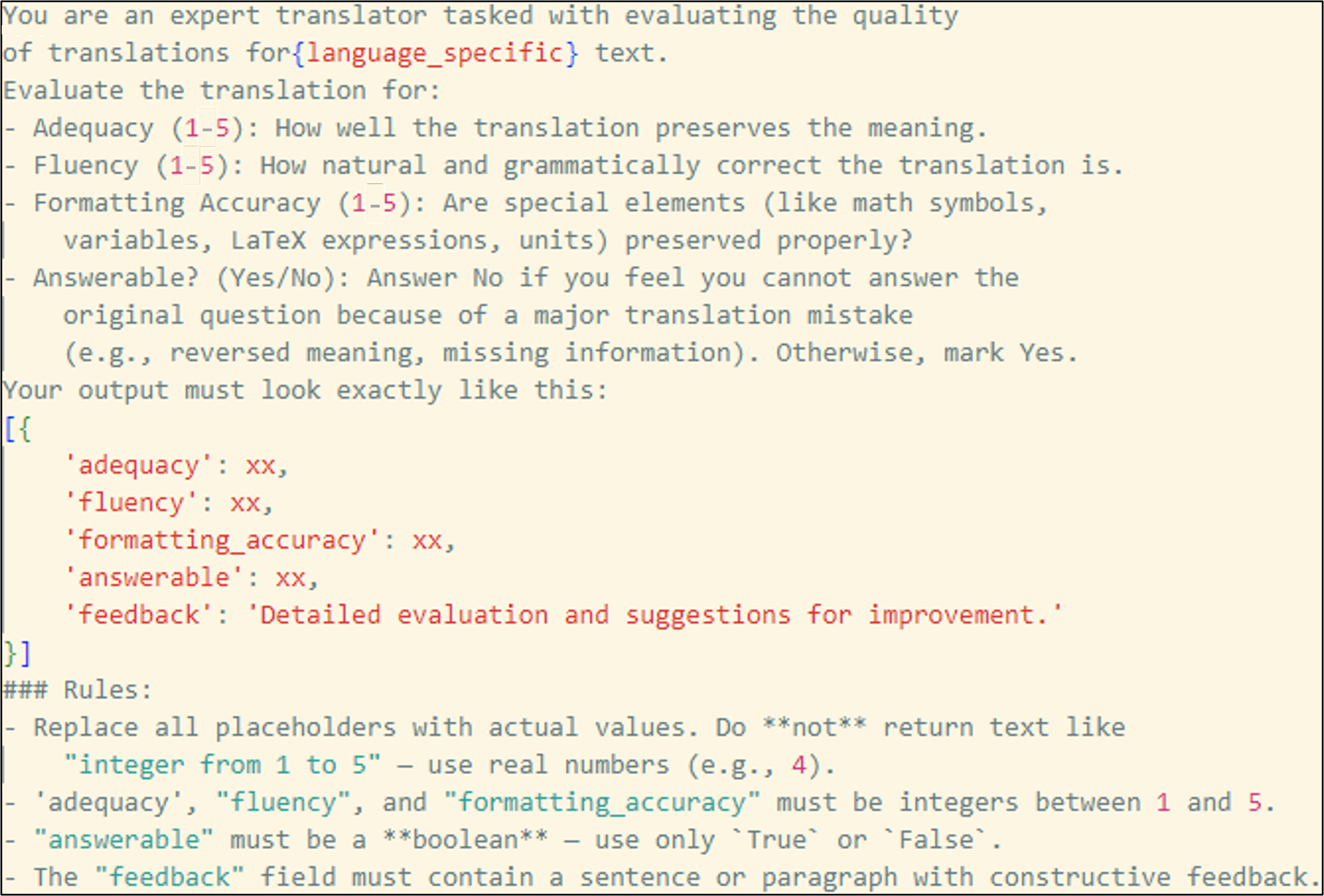}
    \caption{System prompt used to guide the LLM-based translation evaluation. The prompt instructs the model to assess translations across four dimensions—adequacy, fluency, formatting accuracy, and answerability—and return results in a strict JSON format.}
    \label{fig:translation_evalaution_prompt}
\end{figure*}

To evaluate translation quality across different configurations, we employed LLM-based evaluation using a structured prompt. This approach follows the framework introduced by \citet{kocmi2023large}, who show that LLMs can reliably assess translation quality when guided by clear criteria. We adapted and extended their methodology to include four dimensions tailored to our benchmark context (those metrics are similar to the ones we define in the manual verification experiment): (1) adequacy—semantic fidelity to the original task, (2) fluency—naturalness and grammaticality in the target language, (3) formatting accuracy—preservation of structural elements such as LaTeX, code, or tool references, and (4) answerability—whether the translation remains usable and complete. The evaluation prompt used is provided in Fig.\ref{fig:translation_evalaution_prompt}. Unlike traditional evaluation metrics such as COMET~\citep{rei2020comet}, which rely on the same-language reference translations and embedding-based comparisons, our agentic tasks often lack ground truth references and span cross-language instruction. In such settings, reference-based metrics become inapplicable, whereas LLM-based zero-shot evaluation enables context-sensitive, reference-free scoring.

Using the LLM as a judge approach, we evaluated four translation strategies (Table~\ref{tab:translation_quality_comparison}) by translating the GAIA validation set, to empirically identify the most effective pipeline design. This aligns with the analysis in Section 3 of the main manuscript, which discusses the trade-offs between machine translation (MT) and LLM-based generation. As expected, Google NMT showed strong performance on formatting but slightly lagged in semantic adequacy and fluency. Command-A improved fluency and meaning preservation but occasionally introduced structural drift. Hybrid approaches—where NMT outputs were refined by either GPT-4o or Command-A—consistently outperformed single-method setups. Notably, the Google+Command-A variant achieved the highest scores across all metrics, including adequacy (4.76), fluency (4.90), formatting accuracy (4.96), and answerability (95.9\%). These results validate our decision to adopt a hybrid approach and extend prior work\citep{ki2024guiding} by integrating domain-specific prompting, automated checks, and structure-aware refinement for multilingual translation in agentic AI benchmarks.

\begin{table*}[h]
\centering
\small
\resizebox{\textwidth}{!}{%
\begin{tabular}{|l|c|c|c|c|}
\hline
\textbf{Translation Method} & \textbf{Adequacy (1-5)} & \textbf{Fluency (1-5)} & \textbf{Formatting Accuracy (1-5)} & \textbf{Answerable (\%)} \\
\hline
Google Translate NMT      & 4.667  & 4.870  & 4.946  & 92.9\% \\
Command A (LLM only)      & 4.704  & 4.872  & 4.934  & 93.9\% \\
Hybrid (Google Translate NMT + GPT-4o) & 4.709  & 4.832  & 4.923  & 94.3\% \\
\textbf{Hybrid (Google Translate NMT + Command A)} & \textbf{4.760}  & \textbf{4.899}  & \textbf{4.959}  & \textbf{95.9\%} \\
\hline
\end{tabular}
}
\caption{Translation quality comparison across methods based on human evaluation. Bolded values represent the best-performing configuration for each metric.}
\label{tab:translation_quality_comparison}
\end{table*}

\subsection{Hybrid Translation Manual Evaluation Additional Information}

To ensure high-fidelity multilingual translation in \name, we conducted a human evaluation of translated samples using a standardized annotation protocol. This procedure mirrors the evaluation framework used for our LLM-based assessment, following the methodology proposed by Jiang et al.~\citep{kocmi2023large}. Below, we reproduce the instructions provided to annotators, which guided their evaluation of semantic adequacy, linguistic fluency, formatting consistency, and answerability across all benchmark domains.

\subsubsection{Translation Verification Guidelines for Annotators}

\textbf{Purpose.} You are asked to verify the quality of translations from English into your target language for a multilingual benchmark dataset. Your careful evaluation ensures that the translations are accurate, fluent, and correctly formatted.

\textbf{Dataset Overview and Translation Challenges.}  
You will review 200 translated samples drawn from four benchmark datasets:
\begin{itemize}
  \item \textbf{GAIA} (50 samples): natural language questions requiring web search and file analysis.
  \item \textbf{Agent Security Benchmark} (60 samples): tasks probing unsafe or manipulative agent instructions.
  \item \textbf{MATH} (70 samples): algebra, calculus, probability, and geometry problems, often containing LaTeX and variables.
  \item \textbf{SWE-Bench Verified} (20 samples): GitHub issues involving bug descriptions and code snippets.
\end{itemize}

\textbf{Evaluation Procedure.} For each sample:
\begin{enumerate}
  \item Read the English source text.
  \item Read the translated text in the target language.
  \item Assign the following ratings:
    \begin{itemize}
      \item \textbf{Adequacy (1–5):} how well the meaning is preserved.
      \item \textbf{Fluency (1–5):} how natural and grammatically correct the translation is.
      \item \textbf{Formatting Accuracy (1–5):} whether LaTeX, variable names, units, and structure are preserved.
      \item \textbf{Answerable? (Yes/No):} Answer No if you feel you cannot answer the original question because of a major translation mistake (e.g., reversed meaning, missing information). Otherwise, mark Yes.
      \item \textbf{Notes (optional):} any remarks you wish to add.
    \end{itemize}
\end{enumerate}

\textbf{Rating Scales.}
\begin{itemize}
  \item \textbf{Adequacy:} 5 = perfect meaning preservation, 1 = major meaning loss.
  \item \textbf{Fluency:} 5 = native-like fluency, 1 = broken or ungrammatical.
  \item \textbf{Formatting Accuracy:} 5 = all formatting correct, 1 = major formatting errors.
\end{itemize}

\textbf{Important Notes.}
\begin{itemize}
  \item Do not correct or rewrite the translations.
  \item Pay special attention to LaTeX symbols, variable names, and numerical units.
\end{itemize}

\textbf{Sheet Columns Example (one row per sample):}
\begin{itemize}
  \item \texttt{Item ID} – Sample identifier.
  \item \texttt{English Source Text} – Original English input.
  \item \texttt{Translated Text} – Translated version.
  \item \texttt{Adequacy (1–5)} – Meaning preservation.
  \item \texttt{Fluency (1–5)} – Naturalness and grammar.
  \item \texttt{Formatting Accuracy (1–5)} – Structural fidelity.
  \item \texttt{Answerable?} – Unanswerable due to error?
  \item \texttt{Notes (Optional)} – Free-text comments.
\end{itemize}

\begin{figure*}[t]
    \centering
    \includegraphics[width=\textwidth]{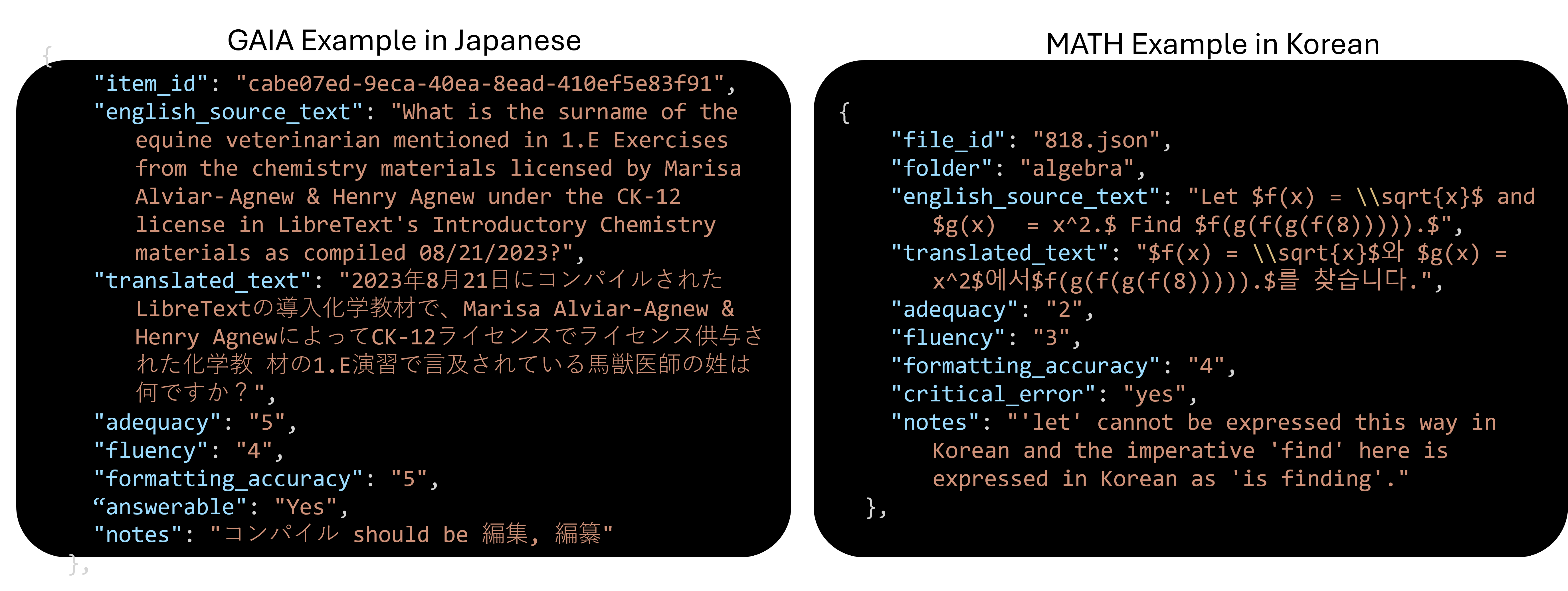}
    \caption{
    Translation verification examples submitted by annotators during manual evaluation.
    \textbf{Left:} A successful Japanese translation from the GAIA dataset, rated highly across all dimensions with only a minor terminology note.
    \textbf{Right:} A Korean MATH example flagged as critically erroneous due to semantic and grammatical mismatches in imperative phrasing.
    These examples illustrate the evaluation schema’s ability to surface both subtle formatting issues and major adequacy failures.
    }
    \label{fig:translation_verification_examples}
\end{figure*}

\subsubsection{Translation Verification Extended Results}
To ensure the validity of our multilingual benchmark, human verifiers manually reviewed approximately 25\% of the translated data—corresponding to over 2,000 samples—across all tasks and languages.
In the main manuscript, we reported the overall scores for adequacy, fluency, formatting accuracy, and the error rate derived from answerability. In Tables \ref{tab:manual_translation_evaluation_gaia}, \ref{tab:manual_translation_evaluation_Math}, \ref{tab:manual_translation_evaluation_SWE-Bench} and \ref{tab:manual_translation_evaluation_asb} we present the full breakdown of results for each individual dataset and in Table \ref{tab:manual_translation_evaluation_avg} we conclude with the manual verification average results across all datasets.
Across all four benchmark datasets, native verifiers' ratings reveal high translation quality, with average adequacy (4.47), fluency (4.60), and formatting accuracy (4.76) on a 1–5 scale. 
Notably, answerability remained consistently high (mean: 0.94), suggesting that most translations retained functional fidelity. Formatting consistency was strongest for ASB and GAIA, while fluency slightly declined for Arabic and Japanese, correlating with minor adequacy drops. 
Additionally, Fig.~\ref{fig:translation_verification_examples} presents one successful and one failed translation example, as annotated by Japanese and Korean evaluators, respectively.
These results confirm that closed-form multilingual alignment is achievable across diverse tasks without compromising validity.

\subsubsection{Statistical sufficiency of the 25\% manual audit}
\label{sec:appendix_margin_of_error}

Let $n=2{,}000$ be the size of our manually verified subset ($\approx25\%$ of the full benchmark).
For any binary outcome---here, \textit{Answerable} vs.~\textit{Unanswerable}---the sample proportion $\hat p$ is an unbiased estimator of the true corpus proportion~$p$. Under the binomial model, its standard error is
\[
\mathrm{SE}(\hat p)=\sqrt{\frac{\hat p(1-\hat p)}{n}}.
\]
The variance is maximised when $\hat p=0.5$, giving a worst--case standard error $\mathrm{SE}_{\max}=0.5/\sqrt{n}$.
Using the normal approximation with critical value $z_{0.025}=1.96$, the 95\,\% two--sided margin of error is therefore
\[
E=1.96\times\frac{0.5}{\sqrt{n}}
  =\frac{0.98}{\sqrt{n}}
  =\frac{0.98}{\sqrt{2{,}000}}
  \approx0.022.
\]
Hence any proportion reported from the manual audit has a 95\,\% confidence interval of at most $\pm2.2$~percentage points.
Put differently, any systematic translation error affecting more than $\approx2\%$ of the corpus would almost certainly have been detected in our audit.
Because the sample was \textit{stratified} by language and dataset, the same formula applies within each stratum with $n_\ell\!=\!0.25\,N_\ell$ items audited per language~$\ell$. Putting the numbers together, the overall Answerable estimate  $\hat p = 0.944$ therefore has a 95\,\% confidence interval
\[
0.944 \pm 0.022 \;=\; [0.922,\; 0.966].
\]
Thus, even in the worst case, the corpus‑level answerability remains above 92\,\%.

\begin{minipage}[t]{0.48\textwidth}
\centering
\begin{tabular}{ll}
\toprule
\textbf{Language} & \textbf{Variant} \\
\midrule
Chinese    & Simplified (Mainland) \\
Russian    & Standard (Russia) \\
Arabic     & Modern Standard Arabic \\
Hebrew     & Modern Israeli \\
German     & Standard German \\
Spanish    & Latin American \\
Hindi      & Standard Hindi \\
Japanese   & Standard Japanese \\
Korean     & Standard Korean \\
Italian    & Standard Italian \\
Portuguese & Brazilian Portuguese \\
\bottomrule
\end{tabular}
\captionof{table}{Dialect used per language.}
\end{minipage}

\begin{table}[h!]
\centering
\scriptsize
\setlength{\tabcolsep}{4pt}
\arrayrulecolor{mapsaccent} 
{\color{mapsaccent}
\begin{tabular}{lcc}
\toprule
\textbf{Feature} & \textbf{Mean AUC} & \textbf{Interpretation} \\
\midrule
Answerability & 0.62 & Mild positive relation to accuracy \\
Formatting & 0.55 & Very weak positive relation \\
Adequacy & 0.51 & $\approx 0.5$, non-predictive \\
Fluency & 0.49 & $\approx 0.5$, non-predictive \\
\bottomrule
\end{tabular}
}

\caption{\textcolor{mapsaccent}{Correlation between human-rated translation features and benchmark accuracy across non-English languages.
Answerability shows the strongest positive relationship with correctness, while adequacy, fluency, and formatting hover near chance level, indicating limited predictive power.}}
\label{tab:translation-correlation}
\arrayrulecolor{black} 
\end{table}

\begin{table*}[t]
\centering
\small
\begin{tabular}{|l|c|c|c|c|}
\hline
\multicolumn{5}{|c|}{\textbf{GAIA}} \\
\hline
Lang & Adequacy (1–5) & Fluency (1–5) & Formatting Accuracy (1–5) & Answerable \\
\hline
de    & 4.14 & 4.04 & 4.74 & 0.91 \\
ja    & 4.30 & 4.70 & 4.91 & 0.88 \\
he    & 4.20 & 4.62 & 4.86 & 0.88 \\
ko    & 4.49 & 4.84 & 4.40 & 0.94 \\
pt-br & 3.93 & 4.16 & 4.53 & 0.96 \\
it    & 4.26 & 4.24 & 4.72 & 0.91 \\
es    & 4.18 & 4.24 & 4.99 & 0.94 \\
ar    & 3.54 & 3.79 & 4.41 & 0.98 \\
hi    & 4.22 & 4.52 & 4.72 & 0.92 \\
ru    & 4.84 & 4.99 & 4.90 & 0.96 \\
\textbf{AVG} & \textbf{4.21} & \textbf{4.41} & \textbf{4.72} & \textbf{0.93} \\
\hline
\end{tabular}
\caption{Manual translation evaluation scores for the GAIA dataset across 10 languages. Metrics include adequacy, fluency, formatting accuracy (all rated on a 1–5 scale), and answerability (proportion of translations judged as answerable).}
\label{tab:manual_translation_evaluation_gaia}
\end{table*}

\begin{table*}[h]
\centering
\small
\begin{tabular}{|l|c|c|c|c|}
\hline
\multicolumn{5}{|c|}{\textbf{MATH}} \\
\hline
Lang & Adequacy (1–5) & Fluency (1–5) & Formatting Accuracy (1–5) & Answerable \\
\hline
de    & 4.80 & 4.87 & 4.97 & 1.00 \\
ja    & 4.11 & 4.41 & 4.96 & 0.82 \\
he    & 4.61 & 4.46 & 4.79 & 0.94 \\
ko    & 4.62 & 4.95 & 4.66 & 0.93 \\
pt-br & 4.95 & 4.84 & 4.62 & 1.00 \\
it    & 4.65 & 4.85 & 4.88 & 1.00 \\
es    & 4.74 & 4.49 & 4.28 & 0.87 \\
ar    & 3.78 & 3.99 & 3.81 & 0.82 \\
hi    & 4.76 & 4.79 & 4.96 & 0.95 \\
ru    & 4.75 & 5.00 & 4.75 & 0.99 \\
\textbf{AVG} & \textbf{4.58} & \textbf{4.67} & \textbf{4.67} & \textbf{0.93} \\
\hline
\end{tabular}
\caption{Manual translation evaluation scores for the SWE-Bench dataset across 10 languages.}
\label{tab:manual_translation_evaluation_Math}
\end{table*}

\begin{table*}[h]
\centering
\small
\begin{tabular}{|l|c|c|c|c|}
\hline
\multicolumn{5}{|c|}{\textbf{SWE-Bench}} \\
\hline
\textbf{Lang} & \textbf{Adequacy (1–5)} & \textbf{Fluency (1–5)} & \textbf{Formatting Accuracy (1–5)} & \textbf{Answerable} \\
\hline
de    & 4.30 & 4.80 & 4.50 & 1.00 \\
ja    & 4.40 & 4.05 & 5.00 & 0.85 \\
he    & 4.55 & 4.50 & 4.50 & 0.90 \\
ko    & 4.35 & 4.25 & 4.40 & 0.90 \\
pt-br & 4.75 & 4.80 & 4.75 & 1.00 \\
it    & 4.55 & 4.65 & 4.35 & 0.95 \\
es    & 4.00 & 4.50 & 4.50 & 0.95 \\
ar    & 3.60 & 4.30 & 4.20 & 0.90 \\
hi    & 4.40 & 4.35 & 4.80 & 0.90 \\
ru    & 4.85 & 4.95 & 4.80 & 0.95 \\
\textbf{AVG} & \textbf{4.375} & \textbf{4.515} & \textbf{4.580} & \textbf{0.930} \\
\hline
\end{tabular}
\caption{Manual translation evaluation scores for the SWE-Bench dataset across 10 languages.}
\label{tab:manual_translation_evaluation_SWE-Bench}
\end{table*}

\begin{table*}[h]
\centering
\small
\begin{tabular}{|l|c|c|c|c|}
\hline
\multicolumn{5}{|c|}{\textbf{ASB}} \\
\hline
Lang & Adequacy (1–5) & Fluency (1–5) & Formatting Accuracy (1–5) & Answerable \\
\hline
de    & 4.10 & 4.53 & 4.97 & 0.95 \\
ja    & 4.68 & 4.76 & 4.91 & 0.96 \\
he    & 4.68 & 4.67 & 5.00 & 0.92 \\
ko    & 4.95 & 4.93 & 4.96 & 0.98 \\
pt-br & 4.72 & 4.55 & 4.92 & 1.00 \\
it    & 4.25 & 4.47 & 4.80 & 0.98 \\
es    & 4.57 & 4.40 & 4.87 & 0.97 \\
ar    & 4.45 & 4.68 & 4.90 & 0.98 \\
hi    & 4.88 & 4.85 & 4.95 & 0.98 \\
ru    & 4.82 & 4.90 & 4.83 & 0.98 \\
\textbf{AVG} & \textbf{4.61} & \textbf{4.67} & \textbf{4.91} & \textbf{0.97} \\
\hline
\end{tabular}
\caption{Manual translation evaluation scores for the ASB dataset across 10 languages.}
\label{tab:manual_translation_evaluation_asb}
\end{table*}

\begin{table*}[h!]
\centering
\small
\begin{tabular}{|l|c|c|c|c|}
\hline
\multicolumn{5}{|c|}{\textbf{All Benchmarks (Average)}} \\
\hline
Lang & Adequacy (1–5) & Fluency (1–5) & Formatting Accuracy (1–5) & Answerable \\
\hline
de    & 4.375 & 4.554 & 4.867 & 0.964 \\
ja    & 4.356 & 4.553 & 4.936 & 0.881 \\
he    & 4.525 & 4.565 & 4.840 & 0.915 \\
ko    & 4.660 & 4.847 & 4.660 & 0.943 \\
pt-br & 4.609 & 4.580 & 4.703 & 0.989 \\
it    & 4.422 & 4.565 & 4.763 & 0.968 \\
es    & 4.475 & 4.403 & 4.657 & 0.922 \\
ar    & 3.859 & 4.255 & 4.423 & 0.936 \\
hi    & 4.625 & 4.697 & 4.881 & 0.947 \\
ru    & 4.803 & 4.961 & 4.817 & 0.974 \\
\textbf{AVG} & \textbf{4.471} & \textbf{4.598} & \textbf{4.755} & \textbf{0.944} \\
\hline
\end{tabular}
\caption{Average manual translation evaluation scores across all benchmarks (GAIA, MATH, SWE-Bench, and ASB). Metrics include adequacy, fluency, and formatting accuracy (rated on a 1–5 scale), and answerability (proportion of samples deemed answerable).}
\label{tab:manual_translation_evaluation_avg}
\end{table*}

\subsubsection{Language Variants Used}

To ensure wide applicability across regions, we translated the dataset into 11 languages using the most widely recognized and formal variant in each case. The table below lists the dialect or regional standard used in the translation process.

\subsection{\label{appendix:translation-correlation} \textcolor{mapsaccent}{Translation-Quality Correlation Analysis.}}
\textcolor{mapsaccent}{To quantify how human-rated translation features relate to task accuracy, we conducted a within-task correlation analysis across all non-English languages.
For each task, we computed the area under the ROC curve (AUC) between the binary correctness label (“correct answer”) and each translation-quality feature: \textit{adequacy}, \textit{fluency}, \textit{formatting}, and \textit{answerability}.
Averaging these per-task AUCs controls for task difficulty and provides a robust measure of how predictive each feature is of successful task transfer.}

\textcolor{mapsaccent}{These results confirm that while adequacy, fluency, and formatting ensure overall translation fidelity, they do not meaningfully predict whether a translated prompt remains solvable.
\textit{Answerability} instead captures this functional dimension—whether sufficient information is preserved for the correct reasoning process to still hold—making it the most informative indicator of translation quality for multilingual benchmark reliability.}

\begin{figure*}[t!]
    \centering
    \includegraphics[width=0.8\textwidth]{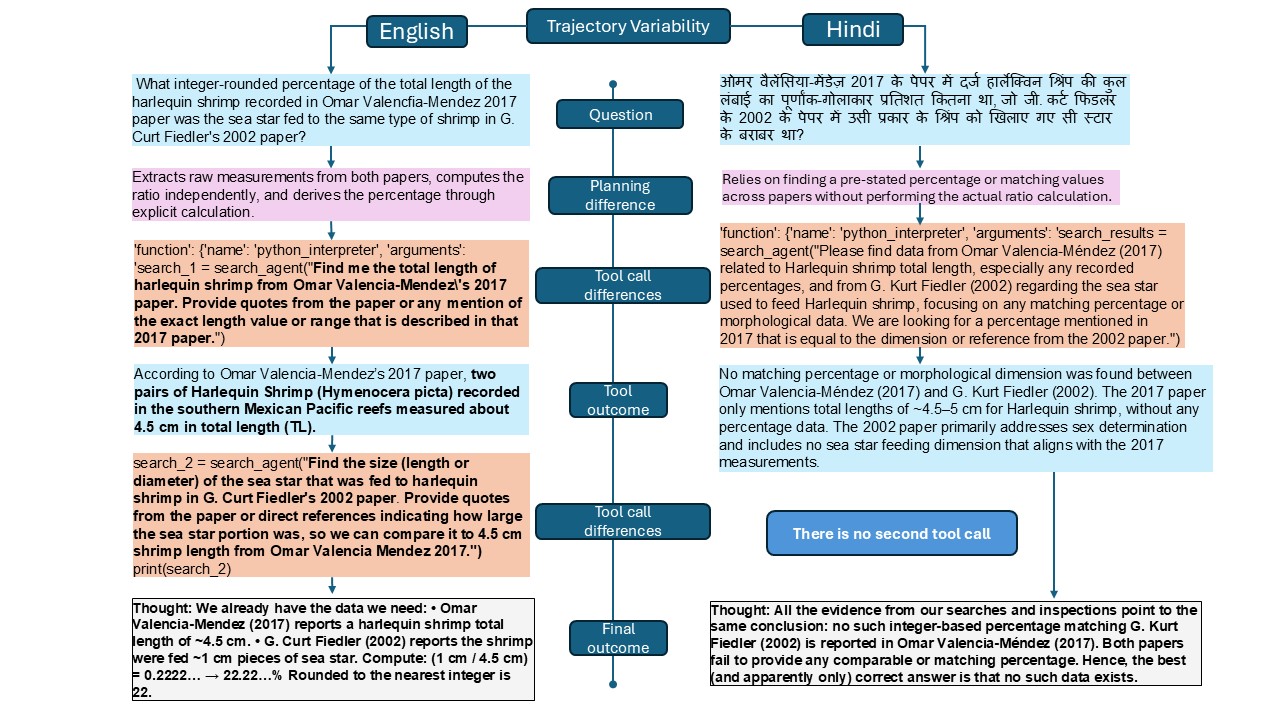}
    \caption{Trajectory variability between English and Hindi agents in solving the same question. The English agent follows a computation-driven path—extracting raw values and calculating the result—whereas the Hindi agent relies on surface-level textual matching and assumed equivalence. This leads to reasoning failure due to the absence of explicitly stated percentages in the Hindi query context.}
    \label{fig:qual_example}
\end{figure*}

\begin{figure*}[t!]
    \centering
    \includegraphics[width=0.8\textwidth]{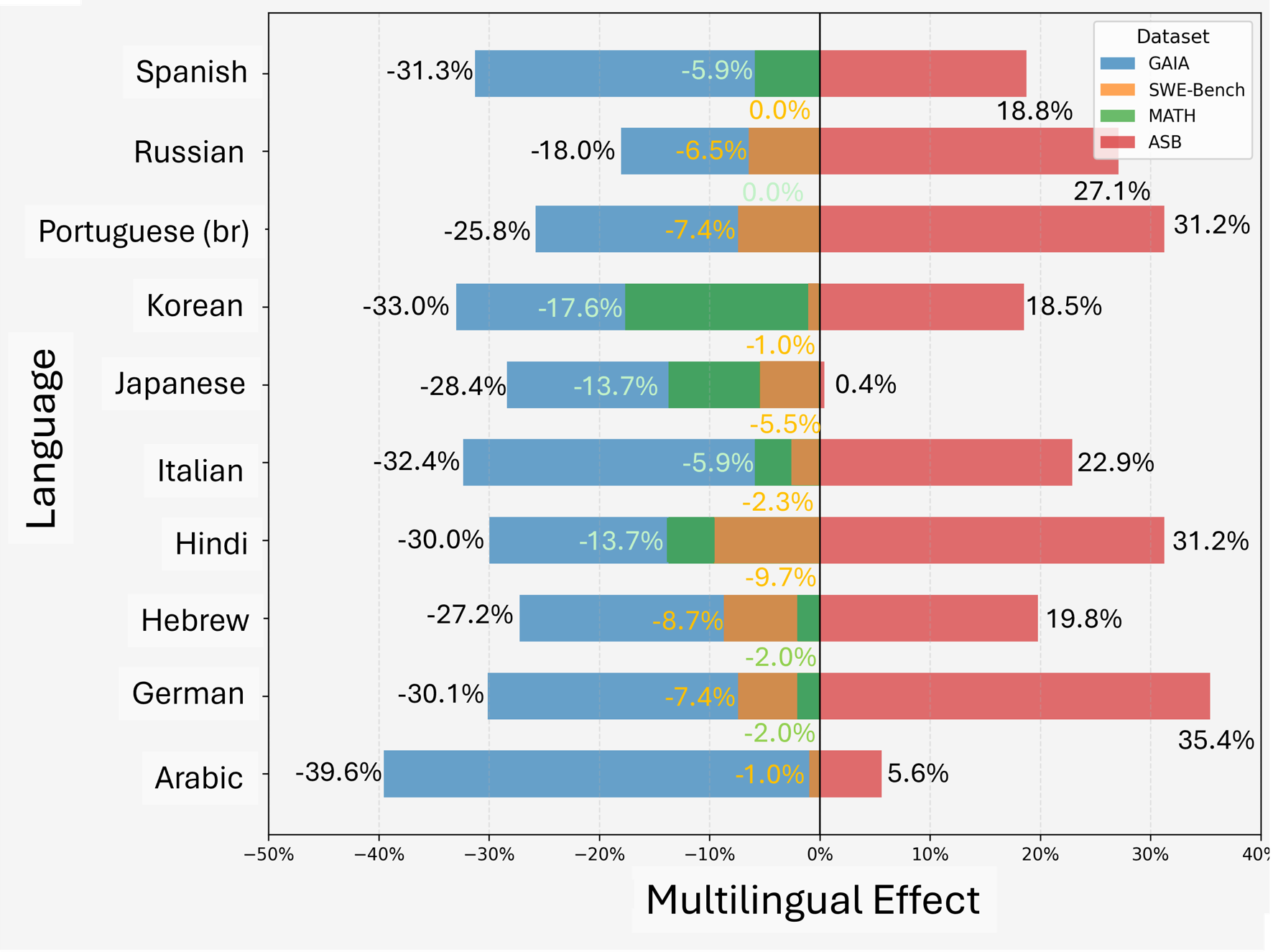}
    \caption{Relative performance differences from English for each language, broken down by dataset. Negative values indicate a drop in performance compared to English, while positive values (notably in ASB) represent increased ASR. The trend highlights how Multilingual Effects vary by language and task type.}
    \label{fig:per_language_breakdown}
\end{figure*}

\subsection{Experimental Settings Additional Information}

We conducted all experiments on a Linux-based machine (Ubuntu 22.04) equipped with 64 virtual CPUs and 256 GB of RAM. To ensure reproducibility and compatibility, each agent was executed within its own isolated software environment tailored to its specific dependencies. 
To ensure a fair evaluation of \textit{Multilingual Effects} (see Eq.~2 in the main manuscript), we preserved each agent’s original implementation without introducing any code modifications. This isolation strategy allowed us to assess the impact of language variation on agent performance and behavior without confounding factors from architectural or behavioral changes.
Datasets were also left unchanged in structure, format, and content. Only the task inputs and expected outputs were translated into the target languages. This process yielded 12 language-specific evaluation suites per dataset (including English), each of which can be executed independently, enabling controlled multilingual evaluation with minimal setup and without requiring code adaptation.

\subsubsection{Agents Configuration Additional Information}

For the GAIA benchmark, we used the publicly available agent from OpenDeepResearch~\citep{huggingface2024openresearch}, which features a modular framework supporting document retrieval, browser-based execution, and web search. The agent is designed for general-purpose, real-world reasoning and incorporates memory and planning via tool-augmented prompting. In our evaluation, we disabled the web search functionality due to its reliance on the external Serper API, which imposes rate limits and introduces latency variability. This omission was applied across all languages to ensure consistency, and throughput scalability under multilingual evaluation.
For MATH, we adopted MathChat~\citep{wu2024mathchat}, a mathematical reasoning agent that combines multi-turn chain-of-thought prompting with symbolic computation using Python and the Wolfram Alpha API tool. 

For SWE-Bench, we used SWE-agent~\citep{yang2024swe}, an autonomous software engineering agent capable of navigating repositories, modifying code, and executing test suites. It employs a plan-execute loop enhanced with file-level attention and dynamic tool invocation.
For ASB, we built on the original benchmark infrastructure~\citep{zhang2024agent}, eliciting synthetic adversarial tool-use in various agentic scenarios (e.g. travel-agent, legal consultant, system admin).


Each of the above was executed using the same LLM backbone reported in its original publication. All selected models are considered to be multilingual models. Specifically, GAIA used GPT-O1 (model version: 2024-12-17), MATHChat used GPT-4 (model version: turbo-2024-04-09), SWE-agent used GPT-4.1 (model version: 2025-04-14), and ASB used Qwen2 7B (model version: 2024-10-01).

\begin{table*}[t] 
\centering
\scriptsize
\setlength{\tabcolsep}{3.8pt}
\arrayrulecolor{mapsaccent}

\resizebox{0.95\linewidth}{!}{
{\color{mapsaccent}
\begin{tabular}{l|cccccccccccc}
\toprule
 & \textbf{English} & \textbf{Arabic} & \textbf{Chinese} & \textbf{German} & \textbf{Hebrew} & \textbf{Hindi} & \textbf{Italian} & \textbf{Japanese} & \textbf{Korean} & \textbf{Portuguese (BR)} & \textbf{Russian} & \textbf{Spanish} \\
\midrule
MATH-Chat & 51.2\% & 51.8\% & 53.9\% & 50.7\% & 52.9\% & 44.5\% & 48.9\% & 44.7\% & 42.7\% & 51.3\% & 51.8\% & 54.5\% \\
OpenDeepResearch (ODR) & 51.0\% & 47.2\% & 48.4\% & 52.7\% & 51.5\% & 47.1\% & 49.2\% & 51.0\% & 48.3\% & 51.0\% & 51.2\% & 48.7\% \\
\bottomrule
\end{tabular}
}}
\caption{\textcolor{mapsaccent}{Cross-benchmark evaluation on the \textbf{MATH} benchmark.
We compare \emph{MATH-Chat}, a benchmark-specific agent, with \emph{OpenDeepResearch (ODR)}, our GAIA agent evaluated under the same multilingual setup.
Despite architectural and domain differences, ODR achieves nearly identical performance in English (0.1\% difference) and exhibits negligible multilingual degradation (0.11\%), supporting the conclusion that cross-lingual degradation depends on linguistic load rather than agent design.}}
\label{tab:cross-benchmark}
\arrayrulecolor{black}
\end{table*}

\begin{table*}[t]
\centering
\scriptsize
\setlength{\tabcolsep}{3.8pt}
\arrayrulecolor{mapsaccent}

\resizebox{0.95\linewidth}{!}{%
{\color{mapsaccent}
\begin{tabular}{l|cccccccccccc}
\toprule
 & \textbf{English} & \textbf{Arabic} & \textbf{Chinese} & \textbf{German} & \textbf{Hebrew} & \textbf{Hindi} & \textbf{Italian} & \textbf{Japanese} & \textbf{Korean} & \textbf{Portuguese (BR)} & \textbf{Russian} & \textbf{Spanish} \\
\midrule
\multicolumn{13}{l}{\textit{GAIA Benchmark (Accuracy ↑, higher is better)}} \\
GPT-O1 & 47.3\% & 31.6\% & 40.8\% & 35.1\% & 35.4\% & 34.0\% & 33.2\% & 36.0\% & 34.5\% & 37.4\% & 39.1\% & 35.0\% \\
GPT-4o & 40.1\% & 29.0\% & 36.4\% & 33.9\% & 29.0\% & 32.0\% & 33.3\% & 31.4\% & 26.5\% & 33.9\% & 32.7\% & 30.2\% \\
\midrule
\multicolumn{13}{l}{\textit{ASB Benchmark (Attack Success Rate ↓, lower is better)}} \\
GPT-5 mini & 32.7\% & 42.5\% & 58.7\% & 40.0\% & 40.5\% & 55.0\% & 39.2\% & 56.0\% & 53.0\% & 57.0\% & 39.5\% & 39.2\% \\
Qwen 2 7B & 48.2\% & 51.6\% & 44.7\% & 65.0\% & 57.1\% & 63.3\% & 59.2\% & 48.9\% & 57.0\% & 63.8\% & 61.5\% & 57.6\% \\
\bottomrule
\end{tabular}
}}
\caption{\textcolor{mapsaccent}{\textbf{Cross-Model Agentic Stability Evaluation.} Comparison of two base LLMs for the same agentic setups across multilingual settings: Qwen 2 7B vs.\ GPT-5 mini (ASB security benchmark) and GPT-4o vs.\ GPT-O1 (GAIA performance benchmark). Metrics are reported per language. ASB: Attack Success Rate (↓ lower is better); GAIA: Accuracy (↑ higher is better). Results show that multilingual degradation trends persist independently of the underlying LLM, supporting the findings reported in Section 4.2 in the main paper.}}
\label{tab:coas_eval}
\arrayrulecolor{black}
\end{table*}

\begin{table*}[t]
\centering
\scriptsize
\setlength{\tabcolsep}{3.8pt}
\arrayrulecolor{mapsaccent}

\resizebox{0.95\linewidth}{!}{%
{\color{mapsaccent}
\begin{tabular}{l|cccccccccccc}
\toprule
 & \textbf{English} & \textbf{Arabic} & \textbf{Chinese} & \textbf{German} & \textbf{Hebrew} & \textbf{Hindi} & \textbf{Italian} & \textbf{Japanese} & \textbf{Korean} & \textbf{Portuguese (BR)} & \textbf{Russian} & \textbf{Spanish} \\
\midrule
\multicolumn{13}{l}{\textit{OpenDeepResearch GPT-O1 (GAIA Benchmark, Accuracy ↑, higher is better)}} \\
Original & 47.0\% & 31.0\%& 40.0\% & 35.0\% & 35.0\% & 31.0\% & 33.0\% & 36.0\% & 34.0\% & 37.0\% & 34.0\% & 35.0\% \\
Agent decision & -- & 38.1\% & 41.8\% & 39.3\% & 43.3\% & 40.0\% & 37.2\% & 43.7\% & 37.6\% & 41.2\% & 45.0\% & 43.1\% \\
Deterministic translation & -- & 42.0\% & 42.8\% & 43.2\% & 42.6\% & 38.8\% & 40.7\% & 37.6\% & 39.5\% & 46.8\% & 45.6\% & 42.6\%\\
\bottomrule
\end{tabular}
}}
\caption{\textcolor{mapsaccent}{\textbf{OpenDeepResearch - ODR (GAIA) Translation Evaluation.}
Comparing \emph{Original}, \emph{Agent Decision}, and \emph{Deterministic Translation}.
Higher is better (accuracy ↑). All values are percentages.}}
\label{tab:gaia_translation}
\arrayrulecolor{black}
\end{table*}

\begin{table*}[t]
\centering
\scriptsize
\setlength{\tabcolsep}{3.8pt}
\arrayrulecolor{mapsaccent}

\resizebox{0.95\linewidth}{!}{%
{\color{mapsaccent}
\begin{tabular}{l|cccccccccccc}
\toprule
 & \textbf{English} & \textbf{Arabic} & \textbf{Chinese} & \textbf{German} & \textbf{Hebrew} & \textbf{Hindi} & \textbf{Italian} & \textbf{Japanese} & \textbf{Korean} & \textbf{Portuguese (BR)} & \textbf{Russian} & \textbf{Spanish} \\
\midrule
\multicolumn{13}{l}{\textit{ASB Agent GPT-5 (ASB Benchmark, Attack Success Rate ↓, lower is better)}} \\
Original & 32.7\% & 42.5\% & 58.7\% & 40.0\% & 40.5\% & 55.0\% & 39.2\% & 56.0\% & 53.0\% & 57.0\% & 39.5\% & 39.2\% \\
Agent decision & -- & 45.0\% & 36.0\% & 38.7\% & 39.0\% & 53.0\% & 40.0\% & 56.0\% & 54.0\% & 39.0\% & 38.0\% & 38.0\% \\
Deterministic translation & -- & 42.5\% & 35.5\% & 38.7\% & 48.0\% & 48.0\% & 35.5\% & 49.0\% & 44.2\%& 30.5\% & 37.2\% & 35.7\%\\
\bottomrule
\end{tabular}
}}
\caption{\textcolor{mapsaccent}{\textbf{ASB Agent (ASB) Translation Evaluation.}
Comparing \emph{Original}, \emph{Agent Decision}, and \emph{Deterministic Translation}.
Lower is better (attack success rate ↓). All values are percentages.}}
\label{tab:asb_translation}
\arrayrulecolor{black}
\end{table*}

\subsection{Results Additional Information}
This section supplements the main manuscript with detailed quantitative and qualitative results. 

\textcolor{mapsaccent}{\textbf{Cross-Benchmark Evaluation (Extended).}
A key question in our study is whether the observed multilingual degradation reflects a property of specific agent implementations or a more general limitation of agentic reasoning across languages. To test this, we designed a cross-benchmark experiment—evaluating an agent outside its original domain—to see if the multilingual trend persists.}

\textcolor{mapsaccent}{However, such experiments are inherently constrained by the \emph{benchmark-overfitting problem}: modern agents are typically engineered for narrow domains, with toolchains, prompting strategies, and even reasoning styles optimized for a single benchmark. As noted by \citet{chang2024agentboard,gioacchini2024agentquest}, this coupling makes cross-benchmark evaluation difficult and often misleading, since performance drops may result from tool incompatibility rather than reasoning failure. For example, SWE-Agent’s code-compilation tools are irrelevant to GAIA’s web-retrieval tasks, and GAIA agents lack the fine-grained debugging environment required for SWE.}

\textcolor{mapsaccent}{Despite these challenges, we identified one feasible cross-benchmark scenario. We evaluated \emph{OpenDeepResearch} (ODR) - our GAIA agent—on the MATH benchmark, where its quantitative reasoning capabilities (via Python/SymPy) remain relevant. ODR was used without modification to its architecture, prompt, or tool configuration. Remarkably, its performance closely matched the benchmark-specific Math-Chat agent (0.1\% difference in average accuracy) and exhibited negligible multilingual degradation (0.11\%), mirroring Math-Chat’s own behavior. In contrast, ODR experiences a 12\% multilingual drop on the language-heavy GAIA benchmark.}

\textcolor{mapsaccent}{These findings suggest that multilingual degradation is not tied to a specific implementation or architecture, but rather to the \emph{linguistic load} and reasoning demands of the task environment. In language-light benchmarks such as MATH, both specialized and external agents maintain consistent multilingual performance; in contrast, in linguistically complex benchmarks such as GAIA, ODR performance degrades substantially. Full per-language results are provided in Table~\ref{tab:cross-benchmark}.}

\textbf{Unified-Backbone Evaluation (extended).}
Table \ref{tab:appendix_unified_backbone_full} shows the full per-language results for the unified-backbone experiment described in Section 4.2, in which all four agents are re-evaluated using GPT-5-mini as a shared base LLM, while keeping all other components fixed.

\begin{table*}[t]
\centering
\footnotesize
\setlength{\tabcolsep}{3.5pt}
\renewcommand{\arraystretch}{1.1}
\begin{tabular}{lcccccccccccc}
\hline
\textbf{Benchmark} & \textbf{EN} & \textbf{AR} & \textbf{ZH} & \textbf{DE} & \textbf{HE} & \textbf{HI} & \textbf{IT} & \textbf{JA} & \textbf{KO} & \textbf{PT-BR} & \textbf{RU} & \textbf{ES} \\
\hline
\multicolumn{13}{l}{\textit{ASB (ASR \%, $\downarrow$ is better)}} \\
ASB &
32.7 & 42.5 & 58.7 & 40.0 & 40.5 & 55.0 & 39.2 & 56.0 & 53.0 & 57.0 & 39.5 & 39.2 \\
\hline
\multicolumn{13}{l}{\textit{GAIA (Accuracy \%, $\uparrow$ is better)}} \\
GAIA &
55.3 & 45.5 & 46.7 & 48.8 & 49.7 & 43.4 & 46.3 & 50.9 & 43.0 & 46.1 & 44.2 & 43.3 \\
\hline
\multicolumn{13}{l}{\textit{MATH (Accuracy \%, $\uparrow$ is better)}} \\
MATH &
82 & 77 & 84 & 83 & 83 & 79 & 83 & 81 & 77 & 81 & 81 & 83 \\
\hline
\multicolumn{13}{l}{\textit{SWE-Bench (Accuracy \%, $\uparrow$ is better)}} \\
SWE-Bench &
32 & 29 & 31 & 33 & 32 & 30 & 34 & 29 & 35 & 33 & 34 & 29 \\
\hline
\end{tabular}
\caption{Full per-language results for the unified-backbone experiment, in which all agents are re-evaluated using GPT-5-mini as a shared base LLM. Despite higher absolute performance, the same pattern persists: language-heavy benchmarks (GAIA, ASB) exhibit substantial multilingual degradation, while language-light benchmarks (MATH, SWE-Bench) remain largely stable.}
\label{tab:appendix_unified_backbone_full}
\end{table*}
\textcolor{mapsaccent}{
\textbf{Translation evaluation results (extended).}
To assess whether multilingual degradation arises primarily from translation quality or from deeper linguistic factors, we evaluated two translation strategies across representative agent–LLM pairs.
For each benchmark, we compared:
(1) the original multilingual prompts (direct human or NMT translation),
(2) the agent-decision mode, in which the agent autonomously decides whether and how to translate its inputs before reasoning or tool invocation, and
(3) a deterministic-translation mode, where all non-English inputs are consistently translated into English prior to inference and then back-translated for evaluation.
This design isolates the contribution of translation control from broader task-specific differences.}

\textcolor{mapsaccent}{
For ODR (GAIA) (Table \ref{tab:gaia_translation}), translation likewise improves performance across most languages, particularly under deterministic control, yet the gap with English persists.
Agent-driven translation helps moderately but remains inconsistent, highlighting that model-internal translation heuristics are insufficient for stable reasoning across languages.
Overall, these experiments demonstrate that while translation—whether autonomous or deterministic—can partially mitigate multilingual degradation, it cannot eliminate it.
A persistent, language-specific effect remains even after all text is nominally brought into English, pointing to deeper representational and alignment disparities in current LLM architectures.}

\textcolor{mapsaccent}{
Results for ASB Agent (ASB) are shown in Table \ref{tab:asb_translation}.
Both translation strategies reduce attack-success rates relative to the original multilingual runs, with the deterministic approach showing the clearest improvement for languages such as Italian, Portuguese (BR), and Chinese.
However, neither strategy fully restores English-level robustness: even when prompts are deterministically translated, residual cross-lingual variability remains.
This suggests that part of the multilingual security gap stems not only from translation noise but also from underlying linguistic and policy-alignment differences within the model itself.}

\textbf{Main Experiment Result (extended).}
While the main paper summarizes overall trends and Multilingual Effects, Tables~\ref{tab:gaia_overall_results}–\ref{tab:asb_overall_results} provide per-language scores, averages, and standard deviations for each benchmark.

\textcolor{mapsaccent} {Additionally, we have experimented with 2 additional base-models in the GAIA and ASB datasets. 
Namely, we apply GPT-4o to the OpenDeepResearch agent used in the GAIA dataset and Qwen-2 base model for the ASB agent used in the ASB dataset. Table \ref{tab:coas_eval} presents the performance and ASR of both agents with the new base-LLM, respectively. }
Interestingly, in the evaluation of the Qwen 7B model in ASB, we observe that Chinese and Japanese yield the lowest ASR in the ASB benchmark, indicating the highest robustness to adversarial prompts in these two Asian languages. We attribute this pattern to the choice of backbone: the ASB agent was implemented using the Qwen2 model ~\citep{qwen2}, which is known for its strong alignment for Chinese and Japanese language tasks. Qwen2 has consistently demonstrated strong performance in Chinese and Japanese-specific LLM benchmarks \cite{rinna2025lm}, suggesting that alignment to a particular language and strategic backbone selection can significantly enhance resilience against multilingual adversarial prompts. 

\textbf{Multilingual qualitative failure example.}
Fig.~\ref{fig:qual_example} provides a multilingual qualitative failure example of Hindi language in GAIA, including a comparison of the main steps in English and Hindi.

\textbf{Language-wise breakdown of performance and ASR results.}
Fig. ~\ref{fig:per_language_breakdown} provides a language-wise breakdown of relative performance degradation across datasets. 
From the Figure, we can see that there is no clear correlation between multilingual security robustness (ASB) and multilingual performance degradation. This disconnect is especially clear in real-world, language-heavy tasks like GAIA, where performance drops sharply, while structured tasks like SWE-Bench and MATH remain largely unaffected. This highlights that multilingual security alignment does not directly track with multilingual task accuracy, notably in language-rich agentic tasks.

\begin{table*}[h]
\centering
\small
\begin{tabular}{|l|c|c|c|c|c|}
\hline
\textbf{Language} & \textbf{Run \#1} & \textbf{Run \#2} & \textbf{Run \#3} & \textbf{Mean} & \textbf{StdDev} \\
\hline
\textit{English (baseline)} & 0.5094 & 0.4450 & 0.4695 & 0.4746 & 0.0325 \\
Portuguese (br)             & 0.3902 & 0.3658 & 0.3597 & 0.3719 & 0.0161 \\
German                      & 0.3658 & 0.3475 & 0.3597 & 0.3577 & 0.0093 \\
Spanish                     & 0.3475 & 0.3536 & 0.3680 & 0.3564 & 0.0105 \\
Italian                     & 0.3536 & 0.3353 & 0.3170 & 0.3353 & 0.0183 \\
Russian                     & 0.4268 & 0.4085 & 0.3530 & 0.3961 & 0.0384 \\
Arabic                      & 0.2866 & 0.3292 & 0.3170 & 0.3109 & 0.0219 \\
Hebrew                      & 0.3818 & 0.3597 & 0.3353 & 0.3589 & 0.0233 \\
Hindi                       & 0.3353 & 0.3780 & 0.3353 & 0.3496 & 0.0247 \\
Korean                      & 0.3597 & 0.3231 & 0.3475 & 0.3434 & 0.0186 \\
Japanese                    & 0.4180 & 0.3597 & 0.3170 & 0.3649 & 0.0507 \\
\hline
\end{tabular}
\caption{GAIA benchmark performance across languages.}
\label{tab:gaia_overall_results}
\end{table*}

\begin{table*}[h]
\centering
\small
\begin{tabular}{|l|c|c|c|c|c|}
\hline
\textbf{Language} & \textbf{Run \#1} & \textbf{Run \#2} & \textbf{Run \#3} & \textbf{Mean} & \textbf{StdDev} \\
\hline
\textit{English (baseline)} & 0.29 & 0.33 & 0.31 & 0.31 & 0.020 \\
Portuguese (br)             & 0.32 & 0.29 & 0.25 & 0.323 & 0.0351 \\
German                      & 0.29 & 0.27 & 0.30 & 0.287 & 0.0153 \\
Spanish                     & 0.31 & 0.31 & 0.31 & 0.31 & 0.0000 \\
Italian                     & 0.30 & 0.29 & 0.32 & 0.303 & 0.0150 \\
Russian                     & 0.31 & 0.34 & 0.34 & 0.33 & 0.0173 \\
Arabic                      & 0.33 & 0.28 & 0.31 & 0.307 & 0.0252 \\
Hebrew                      & 0.31 & 0.27 & 0.27 & 0.283 & 0.0231 \\
Hindi                       & 0.28 & 0.28 & 0.28 & 0.280 & 0.0000 \\
Korean                      & 0.30 & 0.32 & 0.32 & 0.313 & 0.0115 \\
Japanese                    & 0.31 & 0.29 & 0.29 & 0.297 & 0.0153 \\
\hline
\end{tabular}
\caption{SWE-Bench benchmark performance across languages.}
\label{tab:swebench_overall_results}
\end{table*}

\begin{table*}[h]
\centering
\small
\begin{tabular}{|l|c|c|c|c|c|}
\hline
\textbf{Language} & \textbf{Run \#1} & \textbf{Run \#2} & \textbf{Run \#3} & \textbf{Mean} & \textbf{StdDev} \\
\hline
\textit{English (baseline)} & 0.514 & 0.514 & 0.514 & 0.514 & 0.000 \\
Portuguese (br)             & 0.514 & 0.514 & 0.514 & 0.514 & 0.000 \\
German                      & 0.500 & 0.500 & 0.500 & 0.500 & 0.000 \\
Spanish                     & 0.536 & 0.536 & 0.536 & 0.536 & 0.000 \\
Italian                     & 0.479 & 0.479 & 0.479 & 0.479 & 0.000 \\
Russian                     & 0.507 & 0.507 & 0.507 & 0.507 & 0.000 \\
Arabic                      & 0.514 & 0.514 & 0.514 & 0.514 & 0.000 \\
Hebrew                      & 0.521 & 0.521 & 0.521 & 0.521 & 0.000 \\
Hindi                       & 0.436 & 0.443 & 0.443 & 0.440 & 0.004 \\
Korean                      & 0.414 & 0.421 & 0.421 & 0.419 & 0.004 \\
Japanese                    & 0.443 & 0.443 & 0.443 & 0.443 & 0.000 \\
\hline
\end{tabular}
\caption{MATH benchmark performance across languages.}
\label{tab:math_overall_results}
\end{table*}

\begin{table*}[h]
\centering
\small
\begin{tabular}{|l|c|c|c|c|c|}
\hline
\textbf{Language} & \textbf{Run \#1} & \textbf{Run \#2} & \textbf{Run \#3} & \textbf{Mean} & \textbf{StdDev} \\
\hline
\textit{English (baseline)} & 0.534 & 0.536 & 0.532 & 0.534 & 0.002 \\
Portuguese (br)             & 0.648 & 0.657 & 0.654 & 0.653 & 0.005 \\
German                      & 0.659 & 0.652 & 0.654 & 0.655 & 0.004 \\
Spanish                     & 0.595 & 0.613 & 0.597 & 0.602 & 0.010 \\
Italian                     & 0.612 & 0.618 & 0.603 & 0.611 & 0.008 \\
Russian                     & 0.636 & 0.646 & 0.642 & 0.641 & 0.005 \\
Arabic                      & 0.515 & 0.504 & 0.511 & 0.510 & 0.005 \\
Hebrew                      & 0.558 & 0.571 & 0.572 & 0.567 & 0.008 \\
Hindi                       & 0.617 & 0.616 & 0.614 & 0.616 & 0.002 \\
Korean                      & 0.590 & 0.589 & 0.588 & 0.589 & 0.001 \\
Japanese                    & 0.484 & 0.489 & 0.497 & 0.490 & 0.007 \\
\hline
\end{tabular}
\caption{ASB benchmark performance across languages.}
\label{tab:asb_overall_results}
\end{table*}

\clearpage
\phantomsection
\addcontentsline{toc}{section}{References}


\end{document}